\documentclass[12pt,useAMS,usenatbib]{mn2e}

\usepackage{natbib, aas_macros}
\usepackage[dvips]{graphicx}
\usepackage{wrapfig}
\usepackage{color}
\usepackage{amsmath}
\usepackage{amssymb}

\newcommand{\Msun}{M_{\odot}}

\newcommand{\LIR}{L_{\rm IR}}

\newcommand{\fesc}{f_{\rm esc}}

\newcommand{\art}{\rm ART^{2}}
\newcommand{\A}{\rm \AA}
\newcommand{\La}{L_{\alpha}}

\newcommand{\EW}{\rm EW}

\newcommand{\Mpc}{\rm {Mpc}}

\newcommand{\MBH}{{\rm M_{\rm{BH}}}}

\newcommand{\lya}{\rm {Ly{\alpha}}}

\newcommand{\Msunyr}{\rm {M_{\odot}~ yr^{-1}}}
\newcommand{\fescalpha}{f_{\rm esc}^{\lya}}
\newcommand{\fescuv}{f_{\rm esc}^{\rm UV}}

\newcommand{\ergs}{{\rm ergs~s^{-1}}}

\title[$\rm ART^{2}$ : Coupling $\lya$ Line and Multi-wavelength
Continuum]
{$\rm ART^{2}$ : Coupling $\lya$ Line and Multi-wavelength
Continuum Radiative Transfer}
\author[Yajima et al]
{Hidenobu Yajima$^{1, 2}$\thanks{E-mail:yuh19@psu.edu(HY);}, 
Yuexing Li$^{1, 2}$,
Qirong Zhu$^{1, 2}$,
Tom Abel$^{3}$
\\
$^{1}$Department of Astronomy and Astrophysics, Pennsylvania State University,
525 Davey Lab, University Park, PA 16802, USA\\
$^{2}$Institute for Gravitation and the Cosmos, The Pennsylvania State University, University Park, PA 16802\\
$^{3}$Kavli Institute for Particle Astrophysics and Cosmology, SLAC National Accelerator Laboratory, Stanford University, \\
2575 Sand Hill Road, Menlo Park, CA 94025, USA\\
}
 
\begin{document}

\date{Accepted ?; Received ??; in original form ???}

\pagerange{\pageref{firstpage}--\pageref{lastpage}} \pubyear{2008}

\maketitle

\label{firstpage}

%
%
\begin{abstract}

Narrow-band $\lya$ line and broad-band continuum have played important roles in the discovery of high-redshift galaxies in recent years. Hence, it is crucial to study the radiative transfer of both $\lya$ and continuum photons in the context of galaxy formation and evolution in order to understand the nature of distant galaxies. Here, we present a three-dimensional Monte Carlo radiative transfer code, All-wavelength Radiative Transfer with Adaptive Refinement Tree ($\art$), which couples $\lya$ line and multi-wavelength continuum, for the study of panchromatic properties of galaxies and interstellar medium. This code is based on the original version of Li et al., and features three essential modules: continuum emission from X-ray to radio, $\lya$ emission from both recombination and collisional excitation, and ionization of neutral hydrogen. The coupling of these three modules, together with an adaptive refinement grid, enables a self-consistent and accurate calculation of the $\lya$ properties, which depend strongly on the UV continuum, ionization structure, and dust content of the object. Moreover, it efficiently produces multi-wavelength properties, such as the spectral energy distribution and images, for direct comparison with multi-band observations. 

As an example, we apply $\art$ to a cosmological simulation that includes both star formation and black hole growth, and study in detail a sample of massive galaxies at redshifts $z=3.1 - 10.2$. We find that these galaxies are $\lya$ emitters (LAEs), whose $\lya$ emission traces the dense gas region, and that their $\lya$ lines show a shape characteristic of gas inflow. Furthermore, the $\lya$ properties, including photon escape fraction, emergent luminosity, and equivalent width, change with time and environment. Our results suggest that LAEs evolve with redshift, and that early LAEs such as the most distant one detected at $z \sim 8.6$ may be dwarf galaxies with a high star formation rate fueled by infall of cold gas, and a low $\lya$ escape fraction.

\end{abstract}

%
%

%
%
\section{Introduction}
 
The hydrogen $\lya$ emission is one of the strongest lines in the UV band. It provides a unique and powerful tool to search for distant galaxies, as first suggested by \cite{Partridge67}, and evidenced by the successful detection of hundreds of $\lya$ emitting galaxies, or  $\lya$ emitters (LAEs),  at redshifts $z > 3$ over the past decade \citep[e.g.,][]{Hu96, Cowie98, Hu98, Steidel00, Rhoads00, Fynbo01, Hu2002, Fynbo03, Rhoads03, Ouchi03, Kodaira03, Maier03, Hu2004, Dawson04, Malhotra04, Horton04, Taniguchi05, Stern05, Kashikawa06, Shimasaku06, Iye06, Hu2006, Gronwall07, Cuby07, Stark07, Nilsson07, Ouchi08, Willis08, Ota08, Hu2010, Ouchi10}. More recently, \citet{Lehnert10} have discovered the most distant galaxy at $z=8.6$ using the $\lya$ line. 

To obtain a full picture of the physical properties of the high-redshift LAEs, efforts have been made recently to observe these objects in broad-band continuum \citep[e.g.,][]{Gawiser06, Gronwall07, Lai07, Nilsson07, Pitrzkal07, Lai08, Ouchi08, Pentericci09, Ono10A, Ono10B, Hayes10, Finkelstein11, Nilsson11}. One remarkable result from these multi-wavelength surveys is that, LAEs appear to evolve with cosmic time: the LAEs at high redshifts $z >5$ tend to be smaller and younger galaxies, and have less dust extinction and higher equivalent widths (EWs) than their counterparts at lower redshifts $z \sim 3$. 

Despite the rapid progress in the $\lya$ detections, the origin and nature of these distant LAEs, however, remain largely unknown, because the radiative transfer (RT) of $\lya$ line is a complicated process. It depends on the geometry, kinematics, ionization state, and dust content of the surrounding medium, as well as the stellar population and other ionizing sources of the galaxy. To date, there have been a number of theoretical studies on the resonant scattering of $\lya$ photons, analytically in simple geometry \citep[e.g.,][]{Hummer62, Adams72, Harrington73, Harrington74, Neufeld90, Neufeld91, Loeb99}, and numerically \citep[e.g.,][]{Auer68, Avery68, Ahn00, Ahn01, Ahn02, Zheng02, Dijkstra06, Hansen06, Tasitsiomi06, Verhamme06, Laursen07, Laursen09a, Pierleoni09, Faucher10, Zheng10, Zheng11, Schaerer11}. Among the numerical codes, Monte Carlo method is widely employed to solve the $\lya$ RT, owing to the irregular geometry and inhomogeneous distribution of the interstellar medium (ISM), and the slow convergence of the numerical techniques at high optical depths. By applying $\lya$ RT codes to cosmological simulations, several aspects of the LAEs have been investigated by various groups, including observed properties of LAEs at $z \sim 5.7$ \citep{Zheng10, Zheng11}, escape fraction of $\lya$ photons from high-z galaxies \citep{Laursen09b}, effects of ionization on the emergent $\lya$ spectra \citep{Pierleoni09-2}, and mechanism of high EWs in dusty multi-phase ISM \citep{Hansen06}. However, most of these previous works have focused only on the $\lya$ line, and were therefore limited to some aspect of the $\lya$ properties, such as flux and line profile. 

On the other hand, there has been an array of RT codes in continuum over the last several decades, using either ray-tracing methods \citep[e.g.,][]{Rowan-Robinson80, Efstathiou90, Steinacker03, Folini03, Steinacker06}, or Monte Carlo methods \citep[e.g.,][]{Witt77, Lefevre82, Lefevre83, Whitney92, Witt92, Code95, Lopez95, Lucy99, Wolf99, Bianchi00, Bjorkman01, Whitney03,  Harries04, Jonsson06, Pinte06, Li08, Chakrabarti09}. In particular, \cite{Li08} developed a multi-wavelength RT code that employs radiative equilibrium and an adaptive refinement grid, which reproduced the observed dust properties of the most distant quasars detected at $z \sim 6$ when applied to the hydrodynamic quasar simulations of \cite{Li07}. 

However, most of these developments focused mainly on radiation transport in dusty environments, in particular the absorption and re-emission by dust, of local star-forming regions. 
In order to explain the diverse properties of LAEs and their evolution with redshift, it is critical to couple $\lya$ with continuum, in the context of galaxy formation and evolution. Several aspects of such an approach have been attempted in previous studies. For examples, \cite{Pierleoni09-2} incorporated pre-computed grid of $\lya$ RT into ionization calculations, \citet{Tasitsiomi06} considered a combination of $\lya$ and ionization RT, \citet{Laursen09b} studied the effect of ionization and dust on transfer of $\lya$ photons, while \citet{Schaerer11} included dust and non-ionizing UV continuum in the $\lya$ RT calculations. An ultimate goal would be to tackle the ionization of hydrogen, non-ionizing continuum, interstellar dust, and $\lya$ propagation and scattering simultaneously.

Furthermore, the implementation of $\lya$ emission should include two major generation mechanisms, the recombination of ionizing photons, and collisional excitation of hydrogen gas. In addition, relevant ionizing sources such as stars, active galactic nucleus (AGN), and UV background radiation should be included. Last, but not the least, the RT code should incorporate an adaptive refinement grid in order to cover a large dynamical range and resolve dense gas regions where star formation takes place and prominent $\lya$ emission comes from \citep{Laursen09a-2}.

In this work, we present a new, 3-D Monte Carlo RT code, All-wavelength Radiative Transfer with Adaptive Refinement Tree ($\art$), which couples $\lya$ line and multi-wavelength continuum, for the study of panchromatic properties of galaxies and ISM. 
(Note that the ``coupling'' in this paper means the RT of $\lya$ with other continuum photons, not the coupling between RT and hydrodynamics.) Our 
 code improves over the original continuum-only version of \cite{Li08}, and features three essential modules: continuum emission from X-ray to radio, $\lya$ emission from both recombination and collisional excitation, and ionization of neutral hydrogen. The coupling of these three modules enables a self-consistent and accurate calculation of the $\lya$ and multi-wavelength properties of galaxies, as the equivalent width of the $\lya$ line depends on the UV continuum, and the escape fraction of $\lya$ photons strongly depends on the ionization structure and the dust content of the object. Moreover, the adaptive refinement grid handles arbitrary geometry and efficiently traces inhomogeneous density distribution in galaxies and ISM. Furthermore, this code takes into account radiation from both stars and accreting black holes, so it can be used to study both galaxies and AGNs. $\art$ can produce a number of observables, including spectral energy distribution (SED) from X-ray to radio, multi-band fluxes and images, $\lya$ emission line and its EW, which can be directly compared to real observations. It has a wide range of applications, and can be easily applied to simulations using either grid-based or smoothed particle hydrodynamic (SPH) codes
by converting the simulation snapshot to the grid structure of $\art$.

The paper is organized as follows. In \S2, we describe the $\art$ code, which includes implementation of three modules of Continuum, $\lya$ line, and Ionization of hydrogen, the adaptive refinement grid, and the dust model. In \S3, we present the application of $\art$ to a SPH cosmological simulation, and study in details the $\lya$ emission and multi-band properties from individual galaxies at redshifts z=8.5, 6.2, and 3.1 from the simulation. We discuss in \S4 contribution to $\lya$ emission from AGNs, stars, and excitation cooling from gas accretion in our model, and effects of numerical resolutions on the RT calculations, and summarize in \S5. 

%
%
\section{ART$^2$: All-wavelength Radiative Transfer with Adaptive Refinement Tree}

ART$^2$ is a 3-D, Monte Carlo RT code based on the original version of \cite{Li08}, which included the continuum emission from X-ray to radio, and an adaptive refinement grid. We have added two more modules, $\lya$ emission from both recombination and collisional excitation, and ionization of neutral hydrogen, to the current version and couple them self-consistently. The continuum part was described in details in \cite{Li08}. Here we briefly outline the Continuum procedure, and focus on the new implementations of $\lya$ line and Ionization.

\subsection{Continuum Radiative Transfer}

The Continuum module in $\art$ was developed in \cite{Li08}, which adopted the radiative equilibrium algorithm of \cite{Bjorkman01}. In dusty environments, absorbed radiation energy by dust is re-emitted as thermal emission. The re-emitted spectrum depends on the temperature of the dust, which is assumed to be in thermal equilibrium with the radiation field. The radiative equilibrium is ensured by performing the Monte Carlo transfer iteratively until the dust temperature distribution converges, which can be computationally expensive. To accelerate the calculation, $\art$ uses the ``immediate reemission" scheme \citep{Bjorkman01}, in which the dust temperature is immediately updated on absorption of a photon packet, and the frequency of re-emitted photons are sampled from a spectrum that takes into account the modified temperature. The temperature is determined by the balance between the stacked energy absorbed by dust in the cell and the thermal emission from them. The emitted energy in a cell in the time interval $\Delta t$ is

\begin{eqnarray}
\label{eq:Eem} 
 E_i^{\rm em} &=& 4\pi\Delta t \int dV_i \int \rho \kappa_\nu B_\nu(T) \,d\nu \cr
              &=& 4\pi\Delta t \kappa_{\rm P}(T_i) B(T_i) m_i \;,
\end{eqnarray} 
where $\kappa_{\rm P}=\int\kappa_\nu B_\nu\,d\nu / B$ is the Planck mean
opacity, $B = \sigma T^4 / \pi$ is the frequency integrated Planck
function, and $m_i$ is the dust mass in the cell,
the subscript $i$ indicates the $i$'th cell.

Solving the balance between the absorbed and emitted energy, we obtain the dust temperature as follows after absorbing $N_{i}$ packets,

\begin{equation}
    \sigma T_i^4= { {N_i L }\over {4 N_\gamma \kappa_{\rm P}(T_i) m_i} } \;,
\label{eq:REtemp} 
\end{equation}
where $N_\gamma$ is the total number of photon packets in the simulation, and $L$ is the total source luminosity. Note that because the dust opacity is temperature-independent, the product $\kappa_{\rm P}(T_i) \sigma T_i^4$ increases monotonically with temperature. Consequently, $T_i$ always increases when the cell absorbs an additional packet.

The added energy to be radiated owing to the temperature increase $\Delta T$
is determined by a temperature-corrected emissivity $\Delta j_\nu$ in the
following approximation when the temperature increase, $\Delta T$, is small:

\begin{equation}
\Delta j_\nu \approx \kappa_\nu \rho \Delta T {{dB_\nu(T)} \over {dT}} \;.
\end{equation}

The re-emitted packets, which comprise the diffuse radiation field, then
continue to be scattered, absorbed, and re-emitted until they finally escape
from the system. This method conserves the total energy exactly, and does not
require any iteration as the emergent SED, $\nu L_\nu = \kappa_\nu B_\nu(T)$,
corresponds to the equilibrium temperature distribution.


\subsection{$\lya$ Line Transfer}

Hydrogen $\lya$ photon corresponds to the transition between the $n=2$ and $n=1$ levels of a hydrogen atom. It is the strongest H{\sc i} transition. The RT of $\lya$ photons is determined by $\lya$ resonant scattering, dust absorption and scattering, and ionization state of the medium. The process is highly complicated in galaxies owing to the complex geometry and gas distribution. In order to accelerate the numerical convergence of the RT process, Monte Carlo method has been commonly used in a number of $\lya$ codes \citep[e.g.,][]{Zheng02, Dijkstra06-2, Tasitsiomi06, Verhamme06-2, Laursen09a-2, Pierleoni09-2, Faucher10}. Our implementation of $\lya$ line transfer adopts the Monte Carlo method, and the major improvements over many of these codes are that, it is coupled with ionization and multi-wavelength continuum which enables a self-consistent and accurate calculation of the $\lya$ properties, and is incorporated with a 3-D adaptive-mesh refinement grid which efficiently handles arbitrary geometry and inhomogeneous density distribution. Moreover, we treat both recombination and collisional excitation for $\lya$ emission.

\subsubsection{Propagation and Scattering of $\lya$ Photons}

The optical depth $\tau_{\nu} (s)$ of a $\lya$ photon 
with frequency $\nu$ traveling a path of length $s$ is determined by
\begin{equation}
\tau_{\nu}(s) = \int_{0}^{s} \int_{-\infty}^{+\infty}
n(V_{\parallel}) \sigma_{\nu} ~dV_{\parallel} dl,
\end{equation}
where $n(u_{\parallel})$ is the number density of neutral hydrogen gas
with parallel velocity component $V_{\parallel}$,
$\sigma_{\nu}$ is the scattering cross section as a function of frequency. 
In the rest frame of the hydrogen atom, $\sigma_{\nu}$ takes the form
\begin{equation}
\sigma_{\rm \nu} = f_{12} \frac{\pi e^{2}}{m_{\rm e}c}
\frac{\Delta\nu_{\rm L} / 2\pi}
{(\nu - \nu_{0})^{2} + (\Delta\nu_{\rm L}/2)^{2}}
\end{equation}
where $f_{12} = 0.4162$ is the $\lya$ oscillator strength, 
$\nu_{0} = 2.466\times10^{15}~\rm Hz~$ is  the line-center frequency,
$\Delta\nu_{\rm L} =  9.936 \times 10^{7}~\rm Hz$
is the natural line width, and the other symbols have their usual meaning.
Assuming a Maxwellian distribution for the thermal velocity of the encountering atoms, the resulting average cross section is
\begin{equation}
\sigma_{\rm x} = f_{12} \frac{\sqrt{\pi}e^{2}}{m_{\rm e}c\Delta \nu_{\rm D}} H(a, x),
\end{equation}
where
\begin{equation}
H(a, x) = \frac{a}{\pi} \int_{-\infty}^{+\infty} \frac{e^{-y^{2}}}{(x-y)^{2} + a^{2}} dy
\end{equation}
is the Voigt function, $\Delta\nu_{\rm D} = [2k_{\rm B}T/(m_{\rm p}c^{2})]^{1/2}\nu_{0}$ is the Doppler width,
$x = (\nu - \nu_{0})/{\Delta \nu_{\rm D}}$ is the relative frequency of 
the incident photon in the laboratory frame,
and $a = \Delta\nu_{\rm L} / (2 \Delta\nu_{\rm D})$ is the relative line width.

Some previous work approximated the Voigt function with a Gaussian fitting in the core and a power law fitting in the wing. However, since this approximation causes a relative error of a few tens per cent at the transit region, we use the analytical fit of \citet{Tasitsiomi06},
\begin{equation}
H(a, x) = q\sqrt{\pi} + e^{- x^{2}},
\end{equation}
where
\begin{equation}
q = \begin{cases}
0 & {\rm for} ~\zeta \leq 0\\
\left(1+\frac{21}{x^{2}}\right)
\frac{a}{\pi(x^{2}+1)}\prod(\zeta)
& {\rm for} ~\zeta > 0, 
\end{cases}
\end{equation}
with $\zeta = (x^{2} - 0.855) / (x^{2} + 3.42)$ and $\Pi(\zeta) = 5.674 \zeta^{4} - 9.207 \zeta^{3} + 4.421 \zeta^{2} + 0.1117 \zeta$.

This approximation fits the Voigt function well for all frequencies, and the relative error is always less than 1 percent above a temperature of 2K.

The optical depth of dust absorption and scattering is estimated as,
\begin{equation}
d\tau = n_{\rm HI} \sigma(\nu) ds + m_{\rm d} \alpha_{\rm d}(\nu) ds
\label{eq:tau}
\end{equation}
where $m_{\rm d}$ is the dust mass and 
$\alpha_{\rm d} = \alpha_{\rm d, abs} + \alpha_{\rm d, sca}$
is the mass opacity coefficient of absorption and scattering.
When the stacked optical depth through passing cells achieves
an optical depth determined by Equation~(\ref{eq:tau}),
the encountering medium is chosen using a random number by comparing with the fraction, 
$n_{\rm HI}\sigma(\nu) / \left( n_{\rm HI}\sigma(\nu) + m_{\rm d} \alpha_{\rm d} (\nu) \right)$.

When $\lya$ photons are scattered by neutral hydrogen atoms,
the frequency in laboratory frame is changed depending on 
the velocity components of the atoms and the direction of the incidence
and the scattering. Then the velocity components of the directions perpendicular to
the incident direction will follow a Gaussian distribution,
and can be generated by a simple Box-Muller method \citep{Press92}.
However, the parallel component of the velocity depends on relative frequency $x$
of the incident photon owing to the resonance nature of the scattering.
The probability distribution of the parallel component is drawn
from
\begin{equation}
f(u_{\parallel}) = \frac{a}{\pi H(a, x)}\frac{e^{-u^{2}_{\parallel}}}{(x - u_{\parallel})^{2} + a^{2}},
\end{equation}
where $u_{\parallel} = V_{\parallel} / v_{\rm th}$ is velocity of the parallel component normalized by
thermal velocity $v_{\rm th}$ (hereafter $u$ means normalized velocity by $v_{\rm th}$).
To follow this distribution, we use the method of \citet{Zheng02}.

When a photon has a frequency $x < x_{\rm cw}$ in the optically-thick cell, 
where $x_{\rm cw}$ is the boundary between the core
and the wing of the Voigt profile, i.e., where 
$e^{-x^{2}}/\sqrt{\pi} = a / \pi x^{2}$,
it cannot travel long distance and is confined in the cell with numerous scattering.
Only when it has a large x from scattering by high velocity atoms can it escape from the cell
and travel long distance. The photon usually experiences many scatterings ($N \gtrsim 10^{3}$)
before it moves into the wing, it is therefore extremely computation costly to trace this process for all photon packets. In order to speed up the calculation, in particular to avoid the huge number of scatterings in the core, we use a core skipping method developed by \citet{Ahn02-2} and follow the procedure of \citet{Dijkstra06-2} and \citet{Laursen09a-2}. It artificially push the photon in the wing by scattering with atoms which have high velocity in the direction perpendicular to the incident direction. The velocity of the perpendicular component $u_{\perp}$
is generated from a truncated Gaussian,
\begin{equation}
\begin{split}
&u_{\perp, 1} = (x^{2}_{\rm crit} - \rm{ln} R_{1})^{1/2}{\rm cos}2\pi R_{2} \\
&u_{\perp, 2} = (x^{2}_{\rm crit} - \rm{ln} R_{1})^{1/2}{\rm sin}2\pi R_{2}. 
\end{split}
\end{equation}
where $R_{1}$ and $R_{2}$ are two univariates, and we use the critical frequency $x_{\rm crit}$ introduced in \citet{Laursen09a-2}, i.e., 
$x_{\rm crit} = 0.02 e^{\xi (ln a\tau_{0})^\chi}$ 
where $(\xi, \chi) = (0.6, 1.2)$ for $a\tau_{0} \leq 60$, or $(\xi, \chi) = (1.4, 0.6)$ for $a\tau_{0} > 60$.
This acceleration scheme can reduce the calculation time by
several orders of magnitude, and can produce a line profile
which agrees well with analytical solutions in a static slab.

The frequency after scattering depends on the direction in which
the photon is scattered. The direction is given by the phase function,
$W(\theta) \propto 1 + \frac{R}{Q} {\rm cos}^{2}\theta$,
where $\theta$ is the angle between the incident direction $n_{\rm i}$
and the outgoing direction $n_{\rm f}$, and $R/Q$ is the degree of 
polarization. The ratio $R/Q$ becomes 3/7 for the $2P_{3/2}$ state
in $x < x_{\rm cw}$ \citep{Hamilton40}, and 1 for the scattering in the wing \citep{Stenflo80}.
The transition from $2P_{1/2}$ in the core, together with the core skipping scheme, results in isotropic scattering.

The difference in the phase function does not affect the $\lya$ properties such as the escape fraction 
$\fesc$, the emergent line profile and luminosity, even if a single phase function is used for all scatterings \citep[e.g.,][]{Laursen09a-2}. Therefore, we simply assume an isotropic scattering for all the scattering process in this work. We should point out, however, that the phase function becomes important in the polarization effect of $\lya$ line \citep[e.g.,][]{Dijkstra08}, and so care must be taken in dealing with that.

In the scattering process, the final frequency in the laboratory frame is then
\begin{equation}
x_{f} = x_{i} - u_{\parallel} + \mathbf{n}_{f}\cdot\mathbf{u} 
- g(1 - \mathbf{n}_{i}\cdot\mathbf{n}_{f})
\end{equation}
where the factor $g = h_{\rm P} \nu_{0}/m_{\rm H}cv_{\rm th}$ takes into account the recoil effect \citep{Field59, Zheng02}  with $h_{\rm P}$ being the Planck constant. We trace the $\lya$ photon packet until it is absorbed by dust or escapes from a galaxy (e.g., when the photons move out of the calculation box typically 10 times larger than the virial radius of a galaxy).

\subsubsection{$\lya$ Emissivity}

$\lya$ emission is generated by two major mechanisms: recombination of ionizing photons and collisional excitation of hydrogen gas. 

\begin{itemize}

\item {\bf Recombination}: Ionizing radiation from stars, AGNs and UVB can ionize
the hydrogen gas in galaxies. The collision by high-temperature gas can also ionize
the hydrogen. The ionized hydrogen atoms then recombine and create $\lya$ photons via
the state transition $\rm 2P \rightarrow 1S$. The $\lya$ emissivity by the recombination is
\begin{equation}
\epsilon^{\rm rec}_{\alpha} = f_{\alpha } \alpha_{\rm B} h \nu_{\rm \alpha} n_{\rm e} n_{\rm HII},
\end{equation}
where $\alpha_{\rm B}$ is the case B recombination coefficient, and $f_{\alpha}$ is the average number of $\lya$ photons produced per case B recombination. Here we use $\alpha_{\rm B}$ derived in \citet{Hui97}.
Since the temperature dependence of $f_{\alpha}$ is not strong, $f_{\alpha} = 0.68$ is assumed everywhere \citep{Osterbrock06}. The product $h\nu_{\alpha}$ is the energy of a $\lya$ photon, 10.2 eV.

\item {\bf Collisional Excitation}:  High temperature electrons can excite the quantum state of hydrogen gas by the collision. Due to the large Einstein A coefficient, the hydrogen gas can occur de-excitation with the $\lya$ emission. The $\lya$ emissivity by the collisional excitation is estimated by
\begin{equation}
\epsilon^{\rm coll}_{\alpha} = C_{\rm Ly \alpha} n_{\rm e} n_{\rm HI},
\end{equation}
where $C_{\rm Ly \alpha}$ is the collisional excitation coefficient,
$C_{\rm Ly\alpha} = 3.7 \times 10^{-17} {\rm exp}(- h\nu_{\alpha}/kT) T^{-1/2}~\rm ergs\; s^{-1}\; cm^{3}$ \citep{Osterbrock06}.
\end{itemize}

Once the ionization structure have been determined (see Section~\ref{sec:ion}), we estimate the intrinsic $\lya$ emissivity in each cell by the sum of above $\lya$ emissivity, $\epsilon_{\alpha} = \epsilon^{\rm rec}_{\alpha} + \epsilon^{\rm coll}_{\alpha}$. The excitation cooling dominates at $T_{\rm gas} \sim 10^{4-5}~\rm K$, but becomes smaller than the 
recombination cooling at $T_{\rm gas} \gtrsim 10^{6}~\rm K$ \citep[e.g.,][]{Faucher10}.

There is a large ambiguity in the estimation of the excitation cooling from multi-component ISM in SPH simulations \citep{Faucher10}. However, in the present work, we find that the power from stars and AGNs is always larger than or comparable with the excitation cooling. Hence, even the largest possible cooling rate is still sub-dominant to the nebular $\lya$ emission 
(i.e., coming from H{\sc ii} regions around young hot stars). In this paper, we estimate the ionization rate and $\lya$ emissivity for each cell by using the mixed physical quantities of multi-component ISM at first, then reduce it by weighting with the mass fraction of cold gas $f_{\rm cold}$, i.e., $\epsilon_{\alpha}^{\rm coll} = \epsilon_{\alpha, 0}^{\rm coll}\times(1-f_{\rm cold})$, where $f_{\rm cold}$ can reach $\sim 0.9$ depending on the situation.

\subsubsection{Test Calculations}

To test our implementation of the $\lya$ RT, we perform some standard tests against analytical solutions, as well as other numerical results in the literature. 

\begin{enumerate}
\item[1.] {Neufield Test}
\end{enumerate}

As the first test, we carry out the RT calculation in a dust-free slab of uniform gas. The uniform slab was analytically studied by \citet{Neufeld90} in the optically thick limit. In this case, the emergent $\lya$ line profile is given by
\begin{equation}
J(\pm \tau_{0}, x) = \frac{\sqrt{6}}{24} \frac{x^{2}}{\sqrt{\pi}a\tau_{0}} \frac{1}{{\rm cosh}[\sqrt{\pi^{3}/54} (x^{3} - x^{3}_{\rm inj})/a\tau_{0}]},
\end{equation}
where $\tau_{0}$ is the optical depth at the line center from mid-plane to the boundary of the slab
and $x_{\rm inj}$ is the injection frequency $x$.

\begin{figure}
\begin{center}
\includegraphics[scale=0.4]{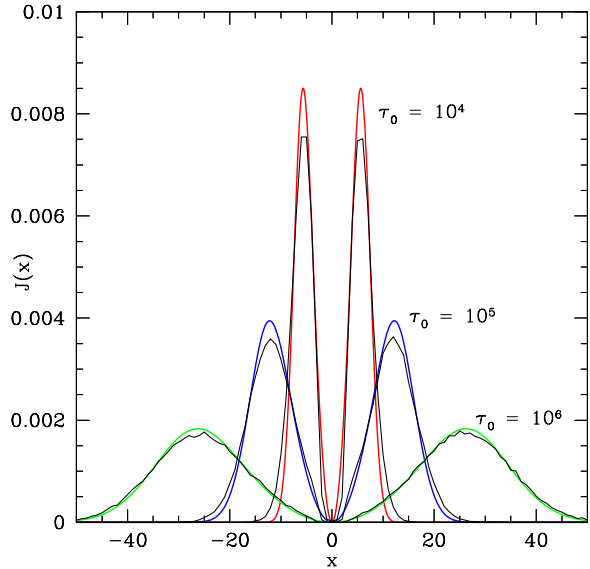}
\caption{
The Neufield Test of emergent $\lya$ profiles of monochromatic line radiation emitted from a dust-free uniform slab. The black lines are our simulation results, and the color lines represent analytical solutions by \citet{Neufeld90} for different optical depths: red -- $\tau_{0}=10^{4}$, blue -- $\tau_{0} = 10^{5}$, and green -- $\tau_{0} = 10^{7}$.  The temperature of gas is 10 K.
}
\label{fig:slab}
\end{center}
\end{figure}

Figure~\ref{fig:slab} shows our test result of emergent spectra from a dust-free gas slab with a temperature of $T = 10 ~\rm K$ and a central plane source, in comparison with analytical solutions for optical depths $\tau_{0} = 10^{4}, 10^{5}$ and $10^{6}$, respectively. The width of the peaks becomes larger with larger $\tau_{0}$, in agreement with the dependence of the optical depth on frequency. Our simulation results agree very well with the analytical solutions, and the agreement becomes better with higher $\tau_{0}$.

The dust absorption is crucial in the study of escape fraction, flux, image and profile of $\lya$ from galaxies. 
\citet{Neufeld90} has derived an approximate expression for the escape fraction of $\lya$ photons in a static dusty slab,
\begin{equation}
f_{\rm esc} = \frac{1}{{\rm cosh}\left[ \zeta^{'}\sqrt{(a\tau_{0})^{1/3}} \tau_{\rm a}\right]},
\end{equation}
where $\tau_{\rm a}$ is the absorption optical depth of dust, and $\zeta^{'} \equiv \sqrt{3}/\zeta \pi^{5/12}$, with $\zeta \simeq 0.525$ being a fitting parameter. This solution is valid for extremely optically thick which $a \tau_{0} \ge 10^{3}$, and in the limit of $(a \tau_{0})^{1/3} \gg \tau_{\rm a}$. It suggests that the escape fraction decreases rapidly with 
increasing $(a\tau_{0})^{1/3}\tau_{a}$.

\begin{figure}
\begin{center}
\includegraphics[scale=0.4]{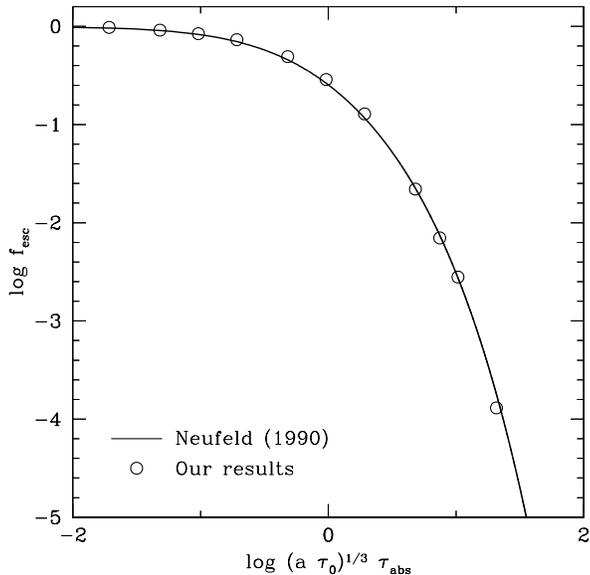}
\caption{
The Neufield Test of escape fraction of $\lya$ photons in a dusty slab as a function of $(a~\tau_{0})^{1/3}~\tau_{\rm abs}$. 
The solid line is an approximate analytical solution of \citet{Neufeld90}, and the open circles are our simulation results.
}
\label{fig:dustslab}
\end{center}
\end{figure}

In figure~\ref{fig:dustslab}, we compare our simulation results of $\lya$ escape fraction in a dusty slab with the analytical solutions of \citet{Neufeld90}. As is shown, our results agree with the analytical curve very well.

\begin{enumerate}
\item[2.] {Loeb \& Rybicki Test}
\end{enumerate}

\citet{Loeb99} analytically derived the intensity field in a spherically symmetric, uniform, radially expanding neutral hydrogen cloud surrounding a central point source of $\lya$ photons. No thermal motion is included ($T = 0$ ~K). In the diffusion limit, the mean intensity $\tilde{J}(\tilde{r}, \tilde{\nu})$ as a function of distance from the source $\tilde{r}$ and
frequency shift $\tilde{\nu}$ is given by
\begin{equation}
\tilde{J} = \frac{1}{4\pi} \left(
\frac{9}{4\pi \tilde{\nu}^{3}}
\right)^{3/2} {\rm exp}\left( -\frac{9\tilde{r}^{2}}{4\tilde{\nu}^{3}}\right),
\end{equation}
with $\tilde{\nu} = (\nu_{0} - \nu)/ \nu_{\star}$, where
$\nu_{\star}$ is the comoving frequency shift of $\lya$ 
at which the optical depth to infinity is unity, and $\tilde{r} = r / r_{\star}$ is the scaled distance,
where $r_{\star}$ is the physical distance at which the Doppler shift
from the source due to the Hubble-like expansion equals $\nu_{\star}$.

\begin{figure}
\begin{center}
\includegraphics[scale=0.4]{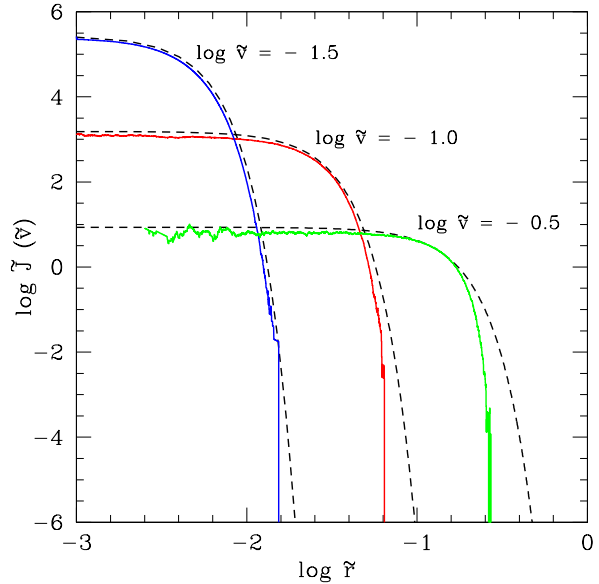}
\caption{
The Loeb \& Rybicki Test of mean intensity profiles of a monochromatic $\lya$ source in a uniformly expanding medium at various frequencies. The dotted lines are analytical solutions by \citet{Loeb99} in the diffusion limit, and the color lines are our simulation results.
}
\label{fig:LR}
\end{center}
\end{figure}

We use a simulation setup similar to that of \citet{Semelin07} for the test. The results are presented in Figure~\ref{fig:LR} in comparison with the analytical solutions of \citet{Loeb99}. Our simulation results are very close to those from previous tests \citep[e.g.,][]{Tasitsiomi06, Semelin07-2}, and are in good agreement with the analytical solutions in regimes where the diffusion limit is valid. However, the simulations diverge from the analytic solutions at larger $\tilde{r}$, where the assumption of optically thick is no longer valid. 

\begin{enumerate}
\item[3.] {Bulk Motion Test}
\end{enumerate}

The bulk motion of gas affects the escape of $\lya$ photons as it decreases the effective optical depth \citep{Dijkstra06-2, Laursen09a-2}. In practice, the bulk speed of the gas to the center can be $\sim 10-1000~\rm km~s^{-1}$ for inflow caused by gravitation, or outflow by supernovae. The relative velocity can decrease the optical depth significantly. Unfortunately, there is no analytical solution of emergent spectrum for moving medium of $T \neq 0$ K. However, \citet{Laursen09a-2} calculated the emergent spectra of an isothermal and homogeneous sphere undergoing isotropic expansion or contraction. Here, we follow the procedure of \citet{Laursen09a-2} to set up the test.  The velocity $\mathbf{v}_{\mathrm{bulk}}(\mathbf{r})$ of a fluid element at a distance $\mathbf{r}$ from the center is set to be
\begin{equation}
\label{eq:vr}
\mathbf{v}_{\mathrm{bulk}}(\mathbf{r}) = \mathcal{H} \mathbf{r},
\end{equation}
where the Hubble-like parameter $\mathcal{H}$ is fixed such that the velocity
increases linearly from 0 in the center to a maximal absolute velocity
$v_{\mathrm{max}}$ at the edge of the sphere ($r = R$):
\begin{equation}
\label{eq:vmax}
\mathcal{H} = \frac{v_{\mathrm{max}}}{R},
\end{equation}
with $v_{\mathrm{max}}$ positive for an expanding sphere, or negative for a collapsing one.

\begin{figure}
\begin{center}
\includegraphics[scale=0.4]{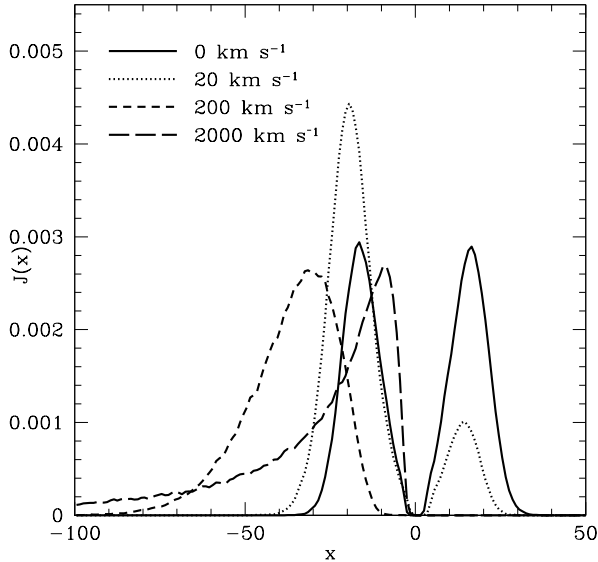}
\caption{
Bulk Motion Test of emergent $\lya$ spectrum from an isothermal and homogeneous sphere undergoing isotropic expansion with different maximal velocities at the edge of the sphere. The initial conditions are the same as those used for Figure 8 in \citet{Laursen09a-2}, with a temperature of the sphere $T = 10^{4}$ K, and the hydrogen column density $N_{\rm HI} = 2\times10^{22}\; \rm cm^{-2}$.  
}
\label{fig:vout}
\end{center}
\end{figure}

Figure~\ref{fig:vout} shows the our test results of the emergent spectrum from an isothermal and homogeneous sphere undergoing isotropic expansion with different maximal velocities $v_{\rm max}$ at the edge of the sphere. As $v_{\rm max}$ increases, the blue wing is suppressed while the red wing is broadened. This is because in an outflowing sphere, the photons in the blue wing have a higher probability of encountering hydrogen atoms and being scattered than those in the red wind, which can escape more easily. This plot suggests that the peak of the profile is pushed further away from the line center for increasing $v_{\rm max}$. However, above a certain threshold value, the peak moves back toward the center again owing to the decrease of
optical depth caused by the fast bulk motion. Our results show a good agreement with the previous simulations by \cite{Laursen09a-2}.



\subsection{Ionization Radiative Transfer}
\label{sec:ion}

Ionizing photons from young stars play an important role in galaxy evolution and the cosmic reionization of the intergalactic medium (IGM) in the early universe. Stellar radiation can ionize interstellar gas, and a fraction of them can escape from galaxies, and reionize the neutral hydrogen \citep[e.g.,][]{Iliev07, Gnedin08a, Razoumov10, Yajima09, Yajima11}. The ionization structure in IGM is highly relevant to the detectability of high-$z$ LAEs \citep[e.g.,][]{McQuinn07, Dayal11}. In particular, ionization affects significantly the properties of distant LAEs, as the emissivity, scattering and escaping of $\lya$ photons depend sensitively on the ionization state of the ISM.

The existing algorithms of ionization RT can be broadly divided into three categories: moments, Monte Carlo, and ray-tracing methods (see \citealt{Trac09} for a recent review). Recently, \cite{Iliev06} and \cite{Iliev09} conducted a comparison of various cosmological ionization codes. It was shown that the ray-tracing method provides the most accurate solution but is computationally expensive \citep[e.g.,][]{Gnedin97, Umemura99, Abel99, Abel02, Razoumov02, Mellema06, Susa06a, Hasegawa10}, the moments method is usually diffusive \citep[e.g.,][]{Gnedin01}, and the Monte Carlo method is efficient and flexible but requires a large number of photon packets \citep[e.g.,][]{Ciardi01, Maselli03, Pierleoni09-2}. 

Our implementation of the ionization RT in $\art$ code is based on the Monte Carlo method, which is similar to CRASH \citep{Maselli03-2}.

\subsubsection{Basic Equations and Methods}

The RT of ionizing photons in our code also includes absorption and scattering process by 
hydrogen gas and interstellar dust. The optical depth is estimated similar to equation
(\ref{eq:tau}), 
$d\tau = n_{\rm HI} \sigma_{\rm H^{0}}(\nu) ds + m_{\rm d} \alpha_{\rm d}(\nu) ds$
, where $\sigma_{\rm H^{0}}(\nu) = 6.3 \times 10^{-18} (\nu/\nu_{\rm L})^{-3}$
is the ionizing cross section of hydrogen with Lyman limit frequency 
$\nu_{\rm L}$ \citep{Osterbrock89}. The time evolution of ionization degree of hydrogen gas is estimated by
\begin{equation}
\begin{split}
n_{\rm H}\frac{dX}{dt} = k^{\gamma}_{\rm star} n_{\rm HI}
+ k^{\gamma}_{\rm UVB} n_{\rm HI}
+ k^{\rm C}n_{\rm HI}n_{\rm e} 
 - \alpha_{\rm rec}n_{\rm HII}n_{\rm e},
\end{split}
\end{equation}
where $k^{\gamma}_{\rm star}$, $k^{\gamma}_{\rm UVB}$ are
the photo-ionization efficiency by star and UVB, respectively,
$k^{\rm C}$ is the collisional-ionization efficiency,
and $\alpha_{\rm rec}$ is the recombination coefficient.
Both $k^{\rm C}$ and $\alpha_{\rm rec}$ are taken from \citet{Cen92}. 
The $k^{\gamma}$ by UVB is estimated by self-shielding model,
in which UVB is optically thin at $n_{\rm H} < n_{\rm crit}$, zero intensity at $n_{\rm H} \geq n_{\rm crit}$.
We set the critical density to be $n_{\rm crit} = 0.0063~\rm cm^{-3}$ based on the observation 
data of column density distribution of neutral hydrogen gas \citep{Nagamine10}.

The photo-ionization rate by stars and AGNs is estimated by absorbed-photon number in a cell $N^{\rm ion}_{\rm abs}$, i.e., $k^{\gamma}_{\rm star}n_{\rm HI} = N^{\rm ion}_{\rm abs}$. We treat the time integration with an adaptive time step 
introduced in \citet{Baek09}, by using three time steps, $\Delta t_{\rm evo}$, $\Delta t_{\rm RT}$ and $\Delta t_{\rm rec}$, where $\Delta t_{\rm evo}$ is an evolution time step to
renew the ionization state. It is determined by splitting
the recombination time scale of volume-mean density by $\sim 100$,
which is close to that of low-density region.
In each $\Delta t_{\rm evo}$, the photon-budget of $N_{\rm ph}$ 
are traced and the ionization rate is estimated.
The evolution of ionization degree is renewed with $\Delta t_{\rm evo}$, 
which is divided by $\Delta t_{\rm rec}$, a time step estimated from recombination time-scale in 
each cell such that the ionization rate is constant.
However, when the ratio of the absorbed photon number
to that of neutral hydrogen in a cell exceeds a pre-set limit ($=0.25$ in this work),
the ionization fraction is renewed with $\Delta t_{\rm RT}$.

In addition, the frequency of the photon is sampled from the source spectrum, similar to that in the continuum RT calculation. We use the stellar population synthesis model of GALAXEV \citep{Bruzual03}
to produce intrinsic SEDs of stars for a grid of metallicity and age, and we use a simple, broken power law for the AGN \citep{Li08}.
A \citet{Salpeter55} initial mass function is used in our calculations.

\subsubsection{Test Calculations}

\begin{figure*}
\begin{center}
\includegraphics[scale=0.83]{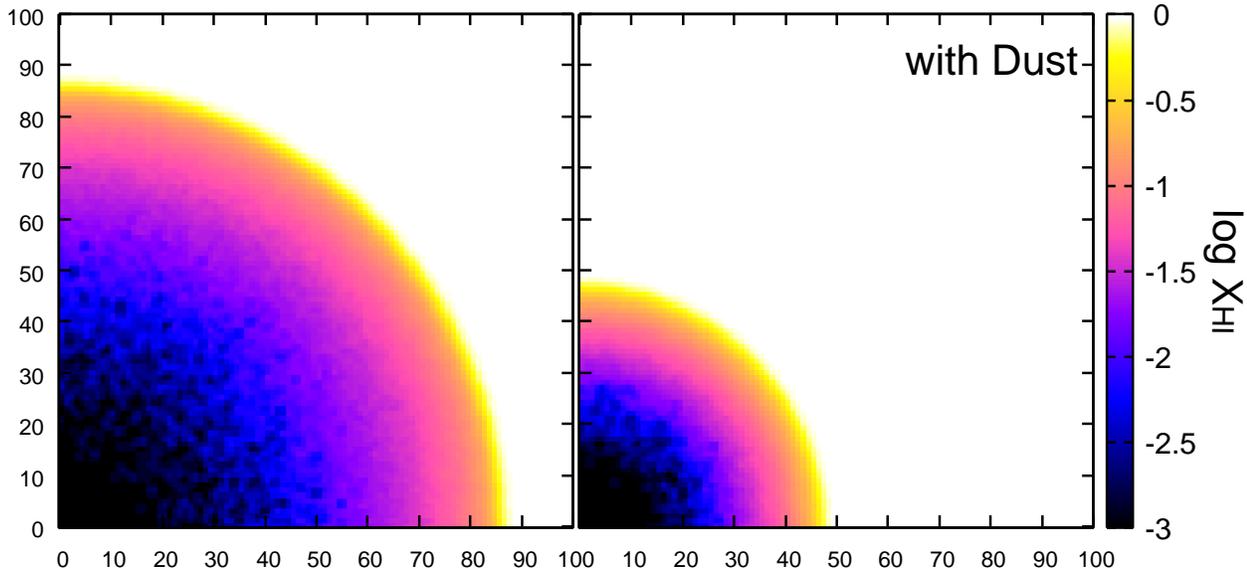}
\caption{
Ionization structure by a point source in a uniform density field. The color bar indicates the neutral fraction of the hydrogen gas in log scale. Left panel: dust-free medium with $n_{\rm H} = 10^{-3}~\rm cm^{-3}$, and right panel: dusty medium with $n_{\rm H} = 10^{-3}~\rm cm^{-3}$ and $\tau_{Sd} = 4.0$, where $\tau_{Sd}$ is a optical depth of dust absorption from source to the radius of the Str\"{o}mgrem sphere. The source is placed at corner of the simulation box with $\dot{N}_{\rm ion} = 5 \times 10^{48}~\rm s^{-1}$, and the cell size is $66.7$ pc.
}
\label{fig:h2img}
\end{center}
\end{figure*}

To test our code, we simulate an H{\sc ii} bubble by a central point source in a hydrogen sphere with uniform density,
i.e., the Str\"{o}mgren sphere \citep{Stromgren39}. The radius of the H{\sc ii} bubble, $r_{\rm s}$, is estimated by the balance between the ionization rate of the source and the recombination rate of the gas
\begin{equation}
\begin{split}
r_{\rm s} = \left( \frac{3 N_{\rm ion}}
{4 \pi \alpha_{\rm B} n_{\rm H}^{2}}
\right)^{\frac{1}{3}},
\end{split}
\end{equation}
where $N_{\rm ion}$ is the number of ionizing photons from the source per second, and $\alpha_{\rm B}$ is the recombination coefficient to all excitation states. 

And the position of the ionizing front of the H{\sc ii} bubble, $r_{\rm i}$ is analytically derived by \citet{Spitzer78},
\begin{equation}
r_{\rm i} = r_{\rm s} \left(
1 - e^{- \frac{t}{t_{\rm rec}}}
\right)^{\frac{1}{3}},
\end{equation}
where $t_{\rm rec}$ is the recombination timescale. When $t  \gg  t_{\rm rec}$, $r_{\rm i}$ approaches to $r_{\rm s}$.

In the presence of dust, the size of H{\sc ii} bubble decreases due to dust extinction.
\citet{Spitzer78} has also derived analytically the size of the H{\sc ii} bubble $r_{i}$ in a dusty medium:
\begin{equation}
3 \int_{0}^{y_{i}} y^{2} e^{y \tau_{Sd}} dy = 1,
\end{equation}
where $y = r/r_{\rm S}$, $y_{i} = r_{i} / r_{\rm S}$, and
$\tau_{Sd}$ is the optical depth of dust from the source to the distance of Str\"{o}mgren sphere.

\begin{figure}
\begin{center}
\includegraphics[scale=0.4]{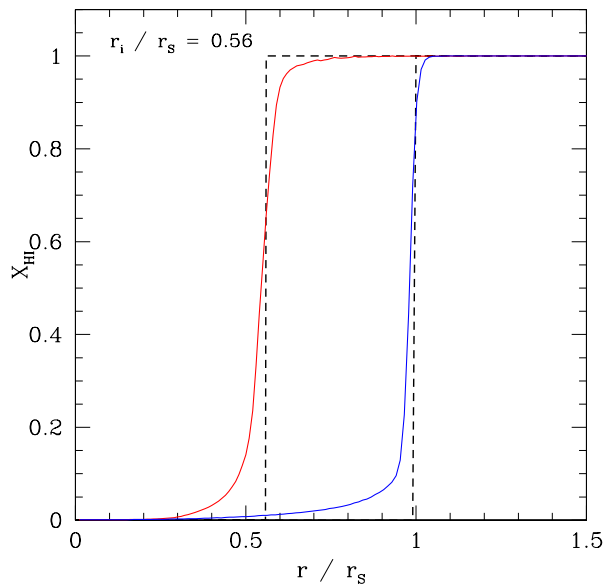}
\caption{
Neutral fraction of hydrogen gas in log scale as a function of distance from source in a dust-free uniform medium.
The distance is normalized by the radius of Str\"{o}mgrem sphere. The color lines represent our simulation results for a dust-free (blue line) and a dusty medium with $\tau_{Sd} = 4.0$ (red line), and the dotted lines are the analytical solutions by \citet{Spitzer78}.
}
\label{fig:x1dist}
\end{center}
\end{figure}

\begin{figure}
\begin{center}
\includegraphics[scale=0.4]{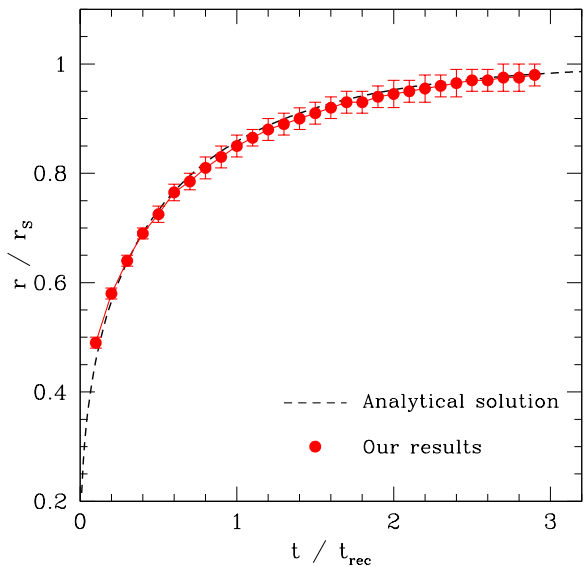}
\caption{
Time evolution of the H{\sc ii} region radius in a dust-free uniform medium. The dotted line is the analytical solution by \citet{Spitzer78}, and the red filled circles are our simulation results. The error bars indicate the radius of neutral fraction of $0.3 - 0.7$.
}
\label{fig:iprop}
\end{center}
\end{figure}

In the tests, we calculate the ionization structure of a uniform density field. The source is placed at the corner of the simulation box with $N_{\rm ion} = 5 \times 10^{48}~\rm s^{-1}$. The number density is $n_{\rm H} = 10^{-3}~ \rm cm^{-3}$, and the recombination coefficient $\alpha_{\rm B}=2.59\times10^{-13}~\rm cm^{3}~s^{-1}$ ($T = 10^{4}~\rm K$). 

Figure~\ref{fig:h2img} shows the neutral fraction of hydrogen gas at mid-plane at $t = 3 t_{\rm rec}$, in a dust-free (left panel) and a dusty medium (right panel). 
The H{\sc ii} bubble becomes significantly smaller than that in dust-free medium. Figure~\ref{fig:x1dist} shows the corresponding neutral fraction of hydrogen gas as a function of the distance from the source for both dust-free and dusty tests, and Figure~\ref{fig:iprop} shows the position of the ionization front of the dust-free medium. All our results are in good agreements with the analytical solutions of \citet{Spitzer78}.


\subsection{Adaptive Refinement Grid}

Our RT calculation is done on a grid. In order to handle arbitrary geometry, and cover a large dynamical range while resolving small-scale, high-density gas regions in galaxies and ISM, we use an adaptive-mesh refinement grid in 3-D Cartesian coordinates. The detailed description of the adaptive grid is given in \cite{Li08}. Here we briefly summarize the basic method and procedures.
 
We typically start with a base grid of $4^{3}$ box covering the entire simulation volume.
Each cell is then adaptively refined by dividing it into $2^{3}$ sub-cells. 
The refinement is stopped if a predefined maximum refinement level, RL, is reached, or if the total number of particles in the cell becomes less than a certain threshold (typically set to be 32 for SPH simulations), whichever criterion is satisfied first. The resolution of the finest level is therefore $L_{\rm min} = L_{\rm box} / 2^{\rm RL + 1}$,
where $L_{\rm box}$  is the box length, and RL is again the maximum
refinement level.

The $\art$ code can be easily applied to either grid- or particle-based simulations, once the snapshot is converted to the format of our grid, which is an effective octo-tree. For particle-based simulations, the physical quantities of particles are interpolated onto a grid. For example, the gas properties at the center of each grid cell, such as density, temperature, and metallicity, are calculated using the SPH smoothing kernel of the original simulation. All physical quantities are assumed to be uniform across a single cell.

\begin{figure}
\begin{center}
\includegraphics[scale=0.48]{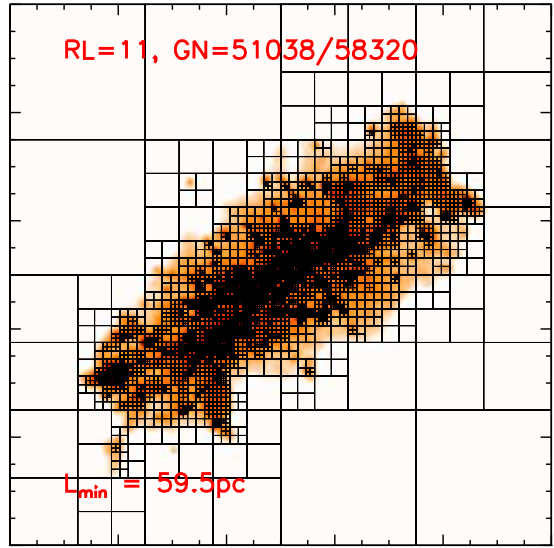}
\caption{
Example of the adaptive grid applied to a snapshot at redshift $z=3.1$ from the cosmological simulation on top of the gas density distribution of the system. The box size is 1 Mpc in comoving scale. The maximum refinement level is 11, and 
the size of the finest cell is 59.5 pc in physical scale.
}
\label{fig:plot_grid}
\end{center}
\end{figure}

Figure~\ref{fig:plot_grid} shows an example of the adaptive grid applied to a snapshot redshift $z=3.1$ from the cosmological simulation used in this work (see Section~\ref{sec_cos} for details). For the parameters used in this application, $L_{\rm box} = 1$ Mpc, RL = 11, we use a total grid number of only $\sim 58300$ and reach a minimum cell size of $L_{\rm min} = 59.5$ pc in physical scale, which corresponds to the spatial resolution of 244 pc in comoving scale of the hydrodynamic simulation.  This grid resolution is equivalent to a $\sim 4096^3$ uniform grid, which is prohibitive with current computational capabilities.


\subsection{ISM Model}

$\art$ implements a multi-phase model of the ISM \citep{McKee77}, as adopted in the hydrodynamic simulations \citep{Springel03a, Springel03b}. It consists of a ``hot-phase'' (diffuse) and a ``cold-phase'' (dense molecular and ${\rm HI}$ core) components, which co-exist under pressure equilibrium but have different mass fractions and volume filling factors (i.e., the hot-phase gas is $\le 0.1$ in mass but $\gtrsim 0.99$ in volume). The cold gas is assumed to follow two empirical relations observed in giant molecular clouds: a power-law mass distribution (e.g., \citealt{Larson81, Fuller92, Ward94, Andre96, Blitz06}), and a power-law mass-radius relation (e.g, \citealt{Sanders85, Dame86, Scoville87, Solomon87, Rosolowsky05, Rosolowsky07}). Within each grid cell in the RT calculation, the hot gas is uniformly distributed, while the cold, dense cores are randomly embedded.

 The dust is assumed to be associated with both the cold and hot phase gases, and it follow the distribution of the gas with certain dust-to-gas ratio. In the RT calculation, when a photon enters a cell, we first determine the distance it travels in the hot phase gas before hitting a cold cloud, then the distance it travels in the cold cloud before it is absorbed. The combination of the two gives the total optical depth in the multi-phase ISM, which is then compared with a random number to determine whether the photon should be stopped for scattering or absorption (see \citealt{Li08} for a detailed description and implementation). 
 The $\lya$ RT in a multi-phase ISM and dust model is one of the new features in our RT code compared with previous works. 
 In $\art$, depending on the hydrodynamic simulation and the nature of the calculation, we can choose either a multi-phase or a single-phase model, in which the gas and the dust are uniformly distributed with the mean density in a cell, for the ISM.

Galaxies and quasars at high redshifts show extinction by interstellar dust
\citep{Ouchi04, Maiolino04}. In the early universe,  most of the dust is thought to be created by Type II supernovae (SNe). Theoretically, several groups have calculated dust formation in the ejecta of Type II SNe \citep{Todini01, Nozawa03, Schneider04, Hirashita05, Dwek07}. In particular, \cite{Todini01} have developed a dust model based on standard nucleation theory and tested it on the well-studied case of SN1987A. They find that SNe with masses in the range of $12-35~\Msun$ produce about 0.01 of the mass in dust per supernova for primordial metallicity, and $\sim 0.03$ for solar metallicity.

In $\art$, we adopt the dust size distribution of \citet{Todini01} for solar metallicity and a $M = 22~\Msun$ SN model, as in Figure $5$ in their paper. The size distribution is then combined with the dust absorption and scattering cross section of \citet{Weingartner01} to calculate dust absorption opacity curves. The SN models include silicate fractions $f_{\rm s} = 0.1$, which shows the silicate feature $\sim 9.7~\rm \mu m$ (see the resultant opacity curve \citet{Li08}). The dust-to-gas mass ratio in each cell is linearly proportional to the metallicity in the cell. The ratio is normalized to Galactic value when the metallicity is the solar abundance.


\subsection{Coupling Ionization, $\lya$ Line, and Continuum}

To date, there is little work on the coupling of continuum and $\lya$ line transfer. In the seminar work of \cite{Pierleoni09-2}, they incorporated pre-computed tables of the properties of the $\lya$ photons in the calculation of ionizing UV continuum radiation. Our approach differs from theirs in that we follow the actual propagation and scattering of the $\lya$ photons in the medium, and take into account the effects of ionization and dust absorption. 

In $\art$, the photons are treated in three regimes: ionizing continuum at $\lambda \le 912\, \A$, $\lya$ line with center $\lambda = 1216\, \A$, and non-ionizing continuum at $ 912\, \A < \lambda \le 10^7\, \A$. We follow the following steps to couple the radiative transfer of these three components:

\begin{enumerate}

\item[1.] The gas and dust content of each grid cell is determined before the RT process.

\item[2.] The neutral hydrogen fraction $n_{\rm HI}$ in each cell is determined by following the ionizing continuum photons.

\item[3.] The resulting ionization structure is then input to calculate the RT of $\lya$ photons.

\item[4.] The absorption of the ionizing photons by dust will be taken into account in the calculation of the dust re-emission in longer wavelengths of the non-ionizing continuum.

\end{enumerate}

\subsection{Monte Carlo Method}

The Monte Carlo RT method follows the propagation, scattering, absorption, and reemission of ``photon packets" (groups of equal-energy, monochromatic photons that travel in the same direction), by randomly sampling the various probability distribution functions that determine the optical depth, scattering angle, and absorption rates. In detail, the Monte Carlo ray-tracing procedure involves the following steps:

\begin{enumerate}

\item[1.] A photon packet of continuum is emitted from either a stellar source or an accreting black hole with random frequencies consistent with the source spectra. In the case of $\lya$ photon, it is emitted from ionized hydrogen gas. The photon is emitted with a uniformly distributed random direction. The probability of a photon being emitted by any given source is determined by its luminosity relative to the total. 

\item[2.] A random optical depth over which the photon must travel before an interaction with gas or dust occurs, $\tau_{i} = - {\rm ln}\xi$, is drawn to determine the interaction location. The interaction includes scattering and absorption. In our method, the photon energies are not weighted, only one event is allowed. That is, at any given interaction site, the photon is either scattered or absorbed, but not both.

\item[3.] Starting from the location of the photon emission, the cumulative optical depth of the photon, $\tau_{\rm tot}$, is calculated stochastically for both hot and cold ISM in a mulit-phase model, or single-phase medium with the mean density along the ray. If the photon is stopped for interaction within a single cell, then $\tau_{\rm tot}$ is the sum of contributions from possibly multiple segments of both hot and cold for the multi-phase model, or a single-phase gas/dust within this cell. If the photon passes through multiple cells before an interaction occurs, then $\tau_{\rm tot}$ is the sum of all contributions from relevant segments in these cells.  

\item[4.] At each boundary between the hot and cold phase gas clouds, or at the boundary of the grid cell, the next interaction point is determined by the comparison between $\tau_{i}$ and $\tau_{\rm tot}$. If $\tau_{i} \leq \tau_{\rm tot}$, then the photon is either scattered or absorbed, with a probability given by the scattering albedo. The exact interaction location is then determined inside either hot or cold phase gas, such that $\tau_{\rm tot}$ becomes exactly $\tau_{i}$. If the photon is scattered, its direction is altered using a phase function, and the ray tracing of the new photon is repeated from step 2. The phase function is the Henyey-Greenstein for dust, and simply isotropic for gas. If the photon is absorbed by dust, the dust temperature is raised, and a new photon is re-emitted according to the scheme described below. The ray tracing of the newly emitted photon again restarts from step 2.

\item[5.] If the photon escapes from the system without reaching the optical depth $\tau_{i}$, it is then collected in the output spectrum and image. The next photon will be picked up from the source, and the whole Monte Carlo procedure from step 1 will be restarted.
\end{enumerate}

\subsection{Making Images}
\label{sec_peeloff}

In $\art$, we incorporate a ``peeling off'' procedure using weighted photons to obtain high-resolution images from a particular viewing direction. From a particular viewing direction, the contribution from direct (radiation source or thermal dust emission) and scattered photons are tracked separately with different weight.  

At each location of emission, the contribution from direct photons to the image is computed as
\begin{equation}
I_{\rm direc}(x,y) = I_0\frac{e^{-\tau_1}}{4\pi d^2},
\end{equation}
where $I_0$ is the energy of each photon packet, $x$ and $y$ are the position of the emission site projected onto the image plane, and $\tau_1$ and $d$ are the total optical depth and distance from the emission site to the observer, respectively. 

At each scattering site, the contribution from scattered photons to the image is 
\begin{equation}
I_{\rm scatt}(x,y) = I_0\frac{e^{-\tau_2}f(\theta)}{d^2},
\end{equation}
where $\tau_2$ is the optical depth from the scattering location to the observer, and $f(\theta)$ is the scattering phase function. 

The total intensity of the image is the sum of $I_{\rm direct}$ and $I_{\rm scatt}$, and is accumulated as each photon packet is emitted and scattered during the course of the Monte Carlo simulation. To obtain an image for a given telescope at a given waveband, a specific filter function is then applied to convolve the image.

%
%

\section{Application to Cosmological Simulations}


\subsection{The Simulations}
\label{sec_cos}

The cosmological simulation presented here follows the formation and evolution of a Milky Way-size galaxy and its substructures. The simulation includes dark matter, gasdynamics, star formation, black hole growth, and feedback processes. The initial condition is originally from the Aquarius Project \citep{Springel08a}, which produced the largest ever particle simulation of a Milky Way-sized dark matter halo. The hydrodynamical initial condition is reconstructed from the original collionsionless one by splitting each original particle into a dark matter and gas particle pair, displaced slightly with respect to each other (at fixed center of mass) for a regular distribution of the mean particle separation, and with a mass ratio corresponding to a baryon fraction of 0.16 \citep{Wadepuhl11}.

 The whole simulation falls in a periodic box of 100 $h^{-1} \Mpc$ on each side with a zoom-in region of a size $5\times 5\times 5~ h^{-3} \Mpc^{3}$. The spatial resolution is $\sim 250~h^{-1}$ pc in the zoom-in region. The mass resolution of this zoom-in region is $\sim 2 \times 10^{5}~ h^{-1} \Msun$ for dark matter particles, $\sim 1.9  \times  10^{4}~ h^{-1} \Msun$ for gas and star particles. The cosmological parameters used in the simulation are $\Omega_{m }= 0.25$, $\Omega_{\Lambda} = 0.75$, $\sigma_{8} = 0.9$ and $h=0.73$, consistent with the five-year results of the WMAP \citep{Komatsu09}. The simulation evolves from $z = 127$ to $z = 0$.

The simulation was performed using the parallel, N-body/Smoothed Particle Hydrodynamics (SPH) code GADGET-3, which is an improved version of that described in \cite{Springel01} and \cite{Springel05e}. For the computation of gravitational forces, the code uses the ``TreePM'' method \citep{Xu95} that combines a ``tree'' algorithm \citep{Barnes86} for short-range forces and a Fourier transform particle-mesh method \citep{Hockney81} for long-range forces. GADGET implements the entropy-conserving formulation of SPH \citep{Springel02} with adaptive particle smoothing, as in \cite{Hernquist89}. 
Radiative cooling and heating processes are calculated assuming collisional ionization equilibrium \citep{Katz96, Dave99}. Star formation is modeled in a multi-phase ISM, with a rate that follows the Schmidt-Kennicutt Law (\citealt{Schmidt59, Kennicutt98}). Feedback from supernovae is captured through a multi-phase model of the ISM by an effective equation of state for star-forming gas \citep{Springel03a}. The UV background model of \cite{Haardt96} is used. 

Black hole growth and feedback are also included in our simulation based on the model of \cite{Springel05d, DiMatteo05}, where black holes are represented by collisionless ``sink'' particles that interact gravitationally with other components and accrete gas from their surroundings. The accretion rate is estimated from the local gas density and sound speed using a spherical Bondi-Hoyle
\citep{Bondi52, Bondi44, Hoyle41} model that is limited by the Eddington rate.
Feedback from black hole accretion is modeled as thermal energy, $\sim 5\%$ of the radiation, injected into surrounding gas isotropically, as described in \cite{Springel05d-2} and \cite{DiMatteo05-2}. This feedback scheme self-regulates the growth of the black hole and has been demonstrated to successfully reproduce many observed properties of local elliptical galaxies (e.g,, \citealt{Springel05a, Hopkins06}) and the most distant quasars at $z \sim 6$ \citep{Li07}. We follow the black hole seeding scheme of \cite{Li07} and \cite{DiMatteo08} in the simulation: a seed black hole of mass $\MBH = 10^{5}~h^{-1} \Msun$ was planted in the gravitational potential minimum of each new halo identified by the friends-of-friends (FOF) group finding algorithms with a total mass greater than $10^{10} h^{-1} \Msun$. 

Each snapshot is processed by an on-flying FOF group finding algorithm with a dark matter linking length less than 20\% of their mean spacing. Other types of particles are then linked to the nearest dark matter particle. A substructure detection algorithm SUBFIND \citep{Dolag09} is then applied to each group to search satellite halos. In the simulation, each FOF group with its substructure is identified as a galaxy (a detailed description of the simulation is presented in Zhu~et~al., in preparation).

In this work, we apply the $\art$ code to the ten most massive galaxies in each snapshot at six redshift bins from $z= 3.1 - 10.2$ and focus on the $\lya$ and multi-band properties. 
In our post-processing procedure, we first calculate the RT of ionizing photons ($\lambda \le 912 \;\rm \AA$) and estimate the ionization fraction of the ISM. The resulting ionization structure is then used to run the $\lya$ RT to derive the emissivity, followed by the calculation of non-ionizing continuum photons ($\lambda > 912 \;\rm \AA$) in each cell. Our fiducial run is done with $N_{\rm ph} = 10^{5}$ photon packets for each ionizing, $\lya$, and non-ionizing components. Because the spatial resolution of the cosmological simulation is not adequate to resolve the multiple phase of the ISM, we assume a single-phase medium in each density grid. The highest refinement of the grid is $\rm RL = 11$.

\subsection{Multi-wavelength Properties of High-redshift Galaxies}

\subsubsection{Flux Images}

\begin{figure*}
\begin{center}
\includegraphics[scale=1.0]{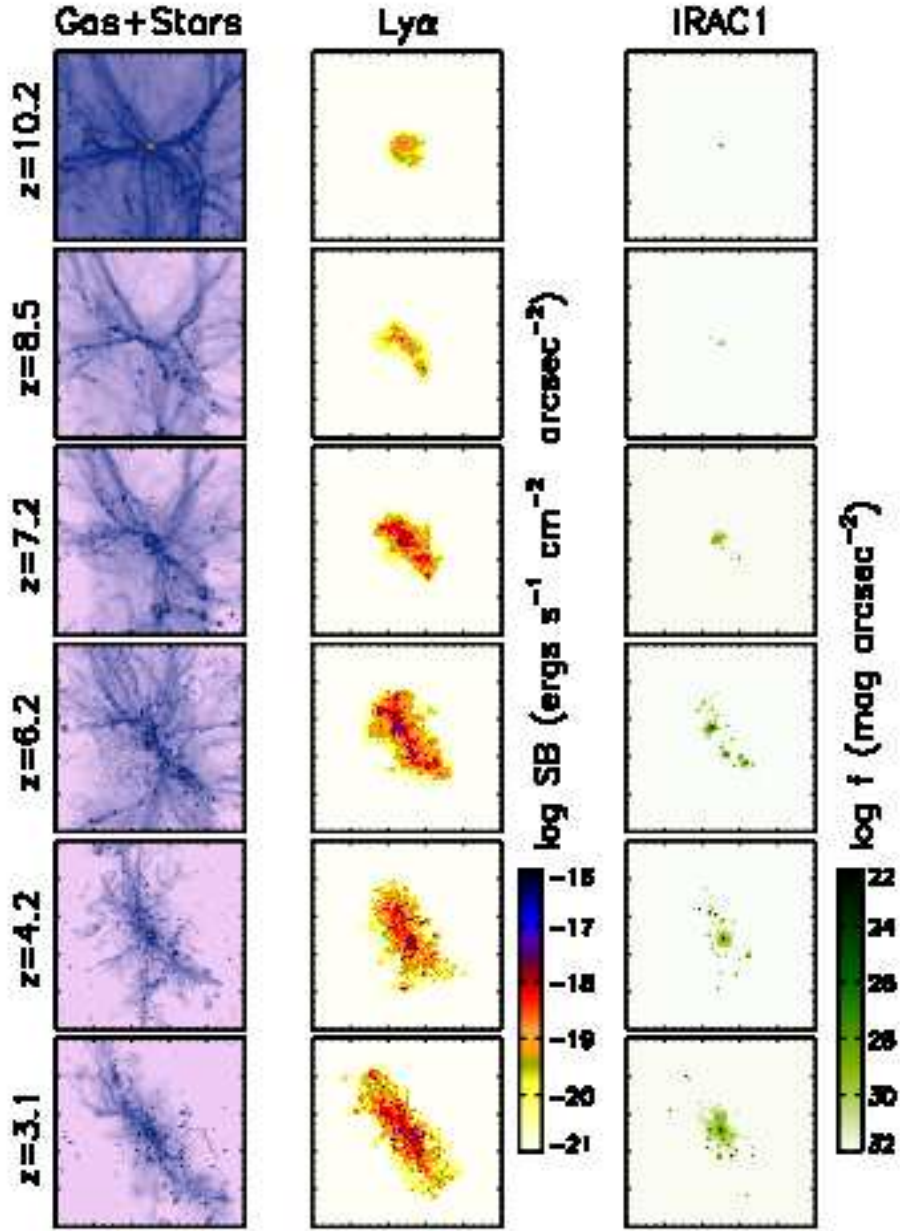}
\caption{
Evolution of galaxies in different wavebands. The left panels show the distribution of  gas and stars from the cosmological simulation. 
Middle panels show the surface brightness of $\lya$, and right 
panels are the flux images in near infrared IRAC-1 ($3.6~\rm\mu m$) on board of $Spitzer$ telescope. 
The box size is 1 Mpc in comoving scale. Note the middle and right panels show only images of the most massive galaxies corresponding to the redshifts labeled in the left column.
}
\label{fig:img_cosmo}
\end{center}
\end{figure*}

The galaxies in our sample are the 10 most massive ones from the cosmological simulation at six snapshots from $z=3.1 - 10.2$. 
Note that because our simulation focuses on a Milky Way-size galaxy, our sample represents the progenitors of the Milky Way at different cosmic time, they are not the typical ``massive'' galaxies one expect from a general cosmological simulation with a mean overdensity. 
The total and stellar masses of these galaxies become larger with time (decreasing redshift) due to gas accretion and merging process.
For example, the total mass of the most massive galaxy is $\sim 2.6\times 10^{10}\, \Msun$, $5.8\times 10^{10}\, \Msun$, $1.6\times 10^{11}\, \Msun$, and $5.4\times 10^{11}\, \Msun$ at $z = 10.2, 8.5, 6.2$ and $3.1$ respectively, 
and the corresponding stellar mass is $\sim 8.5\times 10^{8}\, \Msun$, $4.3\times 10^{9}\, \Msun$, $9.3\times 10^{9}\, \Msun$, and $4.1\times 10^{10}\, \Msun$.

Figure~\ref{fig:img_cosmo} (left panels) show the projected density of gas and stars of each snapshot at the mentioned redshift. 
Filamentary structure is clearly seen, and the most massive galaxy resides in the intersection of the filaments. 
The resulting $\lya$ and near infrared (NIR) fluxes of the most massive galaxy
from our RT calculations are shown in the middle and right panels of Figure~\ref{fig:img_cosmo}, respectively. The NIR band corresponds to IRAC-1 ($3.6~\rm \mu m$) of the {\it Spitzer} telescope.  
The flux image is generated from a ``peeling off'' technique in the RT calculation as described in Section~\ref{sec_peeloff}.

The $\lya$ emission appears to follow the distribution of gas and stars of the galaxy. Its surface brightness and size increase with time (decreasing redshift), while those of the NIR show different pattern. The $\lya$ emission depend on the star formation rate (SFR) of the galaxy and the photon escape fraction. Although SFR decreases at lower redshift, the escape fraction increases as the galaxy grows larger, resulting in an increase of $\lya$ luminosity. In addition, the $\lya$ emission is bright over extended region, owing to many scattering processes. 

On the other hand, the observed IRAC-1 in these redshifts corresponds to wavelengths $\sim 3200 - 9000~\rm \AA$
in the rest frame, hence it depends on both SFR and stellar mass of the galaxy. The cross section of dust decreases steeply with increasing wavelength, it becomes very  small at these wavelengths. Therefore, a significant fraction of photons emitted by the stars in this wavelength range can escape without being scattered or absorbed by dust. Thus, unlike $\lya$, the IRAC-1 image is more compact, and it concentrate on the high density peak of stars. We note that the infrared fluxes are higher than those of the LAE sample at $z=3.1$ by \citet{Gronwall07}, but comparable with those of the IRAC-detected LAEs in \citet{Lai08}. This may be due to high SFR in these galaxies.

The NIR wavelength shown here is similar to the F356W filter of the next generation telescope, the {\it James Webb Space Telescope} (JWST). The detection limit of the filter in AB magnitude can achieve $m_{\rm AB} = 28.1$ with 1 hour exposure time \citep[e.g.,][]{Zackrisson11}. Our calculations show that the model galaxies have $m_{\rm AB} = 24.3~ (z = 3.1), 25.1~ (z = 6.2)$, 28.5 (z = 8.5), and 29.8 (z = 10.2)
at this wavelength. 
Hence, the JWST will be able to detect the infrared emission from galaxies such as the progenitor of the Milky Way at $z \sim 6$ with a $\sim 1$ minute exposure time. 
Even at higher redshifts, these galaxies may be detected with an integration of $\sim 2$ hours for $z = 8.5$, and $\sim 23$ hours for $z = 10.2$.

\subsubsection{Evolution of Spectral Energy Distribution}

\begin{figure}
\begin{center}
\includegraphics[scale=0.4]{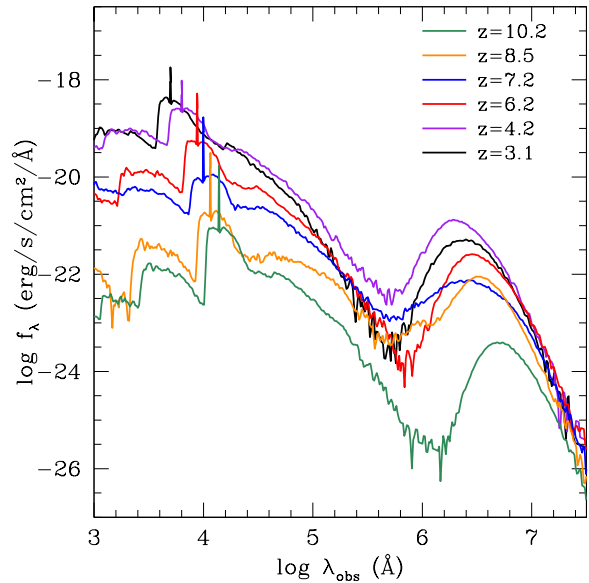}
\caption{
Spectral energy distribution of model galaxies at redshifts $z = 10.2$ (green), 8.5 (orange), 7.2 (blue), 6.2 (purple), and 3.1 (black), respectively.
}
\label{fig:sed_cosmo}
\end{center}
\end{figure}

The corresponding multi-wavelength SEDs of the most massive galaxy at selected redshift
 are shown in Figure~\ref{fig:sed_cosmo}. 
The shape of the SED changes significantly from $z = 10.2$ to $z = 3.1$,
as a result of changes in radiation from stars, absorption of ionizing photons by gas and dust, and re-emission by dust in the infrared. The $\lya$ line appears to be strong in all cases. The deep decline of UV continuum at $z > 8$
 is caused by strong absorption of ionizing photons by the dense gas. Galaxy at lower redshift has a higher floor of continuum emission from stars and accreting BHs, a higher ionization fraction of the gas, and a higher infrared bump owing to increasing amount of dust and absorption. Moreover, due to the effect of negative k-correction, the flux at $\gtrsim 500~\rm \mu m$ stays close in different redshifts. Our calculations show that the model galaxies have a flux of $f_{\nu} =  0.043~ (z = 3.1),~0.057~ (z = 6.2)$,  
$0.02~ (z = 8.5)$, and $0.004~ (z = 10.2)~\rm mJy$
 at $850~\rm \mu m$ in observed frame. The new radio telescope, {\it Atacama Large Millimeter/submillimeter Array} (ALMA) may be able to detect such galaxies at $z \sim 6$ with $\sim 2$ hours integration, and at $z \sim 8.5$ for $\sim 20$ hours with 16 antennas.
Even with ALMA, it may be very difficult to detect such a galaxy at $z = 10.2$ (more than 70 days integration).  

From the SEDs, it is interesting to compare the emergent $\lya$ luminosity and that of the infrared integrated from 8 to 1000 $\rm \mu m$ in the rest frame. 
A simple relation between infrared and $\lya$ luminosity has been suggested by \citet{Kennicutt98} through star formation, 
namely $\rm {SFR} ({\rm \Msunyr}) = 4.5\times 10^{-44}~ \LIR (\ergs)$ and $\rm{SFR} ({\rm \Msunyr}) = 0.91 \times 10^{- 42}~ \La (\ergs)$
with the assumption of $\La / L_{\rm H\alpha} = 8.7$ (case B).  
However, our calculation suggests a more complex relation between $\lya$ and infrared luminosity, as inferred from Figure~\ref{fig:sed_cosmo}. This is because $\La$ depends on $\fescalpha$, while the $\LIR$ depends on $\fescuv$, and the two escape fractions differ from each other. 
In our simulations, the $\fescuv$ mostly decreases with redshift, and is below 0.5 at $z \lesssim 6$. Hence, the UV continuum may serve as a good tracer of SFR at $z \gtrsim 6$, while $z \lesssim 6$ the infrared flux may be one. 
In addition, since the modeled galaxies at $z \lesssim 6$ show a smaller value of $\fescalpha$ than 0.5 (see next subsection), hence $\lya$ flux also may not be a good indicator of SFR.

\subsubsection{The $\lya$ Properties}
\label{sec_lya}

\begin{figure}
\begin{center}
\includegraphics[scale=0.45]{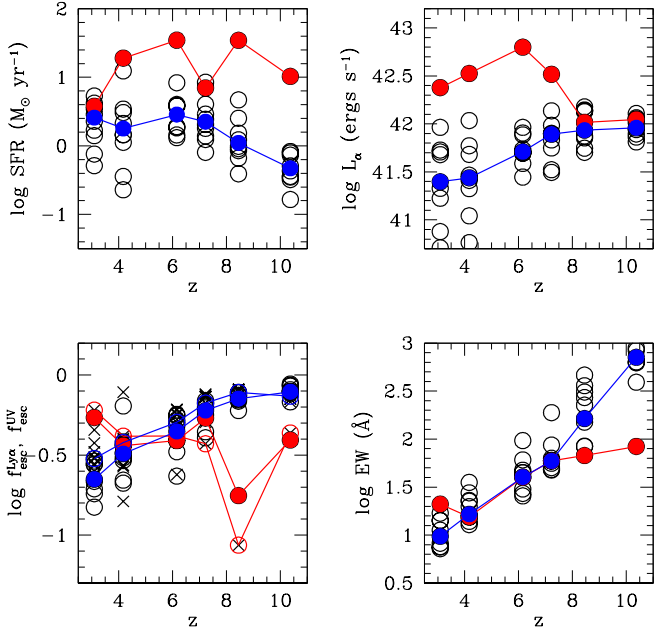}
\caption{$\lya$ properties of modeled galaxies at different redshifts, in clock wise direction: star formation rate, emergent $\lya$ luminosity, equivalent width of $\lya$ line in rest frame, and photon escape fraction of $\lya$ (open circles) and UV continuum (1300 - 1600 $\A$, crosses).
Red and Blue filled circles are the values of the most massive galaxies and the median in our sample at each snapshot, respectively.
Filled and open red circles in the panel of escape fraction shows the $\lya$ and UV continuum escape fraction of the most massive galaxies, 
filled and open blue circles are the median values of $\lya$ and UV continuum, respectively.
}
\label{fig:flux_cosmo}
\end{center}
\end{figure}

The derived properties of $\lya$ emission from these modeled galaxies are shown in Figure~\ref{fig:flux_cosmo}, in comparison with their corresponding SFR. 
The most massive galaxies 
at very high redshift ($z > 6$) maintain a high SFR of $\sim 10-30~\Msunyr$
fueled by accretion of cold gas and merging process. The SFR decreases to $\sim 3 \; \Msunyr$ at $z \sim 3$ owing to depletion of star-forming gas and suppression by radiation feedback from both stars and accreting BHs. The resulting emergent $\lya$ luminosity from recombination of ionizing photons and from excitation cooling increases from $\sim {1.0} \times 10^{42}\, \ergs$ at $z=8.5$ to $\sim {2.3} \times 10^{42}\, \ergs$ at $z=3.1$. 
The observed LAE luminosity functions show a characteristic luminosity $L_{\lya}^{*} \sim 4.4\times 10^{42}\, \ergs$ at $z=6.6$ \citep{Ouchi10}, and $L_{\lya}^{*} \sim 6\times 10^{42}\, \ergs$ at $z=3.1$ \citep{Ouchi08, Gronwall07, Blanc11}, which are comparable to the luminosity range of our modeled galaxies.
The median values of SFR moderately decrease with increasing redshift, and show $\sim 0.5 - 3~\Msunyr$. 
However, the $\La$ does not decreases largely by the redshift evolution of $\fesc$.
In addition, a few galaxies in each snap shot are $\lya$ bright, and have $\lya$ luminosity of $\La \ge 10^{42}~\ergs$, which is detectable with recent observational studies
\citep[e.g.,][]{Ouchi08}.


The lower-left panel of figure~\ref{fig:flux_cosmo} shows the photon escape fraction of $\lya$, $f_{\rm esc}^{\lya}$ and UV continuum $f_{\rm esc}^{\rm UV}$, where $f_{\rm esc}^{\rm UV}$ is calculated at $\lambda_{\rm rest} = 1300 - {1600} \;\A$. Both escape fractions appear to increase with time (decreasing redshift), with $f_{\rm esc}^{\lya} \sim {0.18}$ at $z=8.5$ to  $\sim {0.5}$ at $z=3.1$, while $f_{\rm esc}^{\rm UV} \sim {0.09}$ at $z=8.5$ to  $\sim {0.6}$ at $z=3.1$. 
In the early evolution phase, the most massive galaxies are compact and gas-rich. They grow rapidly from accretion of cold gas which fueled intense star formation. The highly concentrated gas and dust efficiently absorb the $\lya$ and UV photons, and suppress their escape. At a later time, the gas and stars distribute to a more extended region, and a large fraction of gas is converted to stars. Hence, the photon may easily escape from the galaxy, contributing to a higher photon escape fraction.

On the other hand, the median value of $\fesc$ of $\lya$ and UV increases monotonically with redshift, e.g., 
$f_{\rm esc}^{\lya} \sim {0.75}$ at $z=8.5$ to  $\sim {0.29}$ at $z=3.1$, while $f_{\rm esc}^{\rm UV} \sim {0.81}$ at $z=8.5$ to  
$\sim {0.38}$ at $z=3.1$. Such a trend is consistent with observations by \citet{Hayes11}. 

Unlike the most massive galaxies, the low-mass ones have lower SFR and lower dust content, the ISM is less dense and less clumpy, and the metallicity is lower. As a result, the escape fractions are higher in small galaxies than the massive ones.


The EW of $\lya$ line is defined by the ratio between the $\lya$ flux and
the UV flux density $f_{\rm UV}$ in rest frame, where the mean flux density of $\lambda = 1300 - {1600} \; \A$ {in rest frame} is used. In practice, the EW is estimated by $\EW = \EW_{\rm int} \frac{f_{\rm esc}^{{\lya}}}{f_{\rm esc}^{\rm UV}}$, where $\EW_{\rm int}$ is the intrinsic equivalent width.
The $\EW_{\rm int}$ depends on the stellar SED and the excitation $\lya$ cooling: 
the younger SED (or top-heavy IMF) and the more efficient $\lya$ cooling, the higher $\EW_{\rm int}$.

The resulting $\lya$ EWs of the most massive galaxies
are shown in the lower right panel of Figure~\ref{fig:flux_cosmo}. Since the photon escape fraction of  $\lya$ is close to that of the UV continuum, the EW shown is basically $\EW_{\rm int}$. All three modeled galaxies have $\EW > 20\; \A$, they are therefore classified as LAEs \citep[e.g.,][]{Gronwall07}. 
The EW evolves gradually with redshift. At  $z > 6$,  the EWs are high, $\EW \sim {60} \; \A$, owing to strong $\lya$ emission from excitation cooling and AGN activity, while at $z=3.1$, it drops to $\sim 21 \; \A$ as the $\lya$ emission is mainly produced by stellar radiation. 
The median value increases more steeply with redshift, from $\sim 14~\A$ at $z=3.1$ to $\sim 164$ at $z = 8.5$.
These EWs are within the observed range, and the trend is in broad agreement with observations that galaxies at higher redshifts appear to have higher EW than their counterparts at lower redshifts \citep[e.g.,][]{Gronwall07, Ouchi08}.
Some galaxies at higher redshift $z \gtrsim 8$ show very high EW because the excitation $\lya$ cooling becomes dominant \citep{Yajima11b}.
However, it will be suppressed significantly by IGM at such a high redshift.

The recent discovery of the most distant LAE, UDFy-38135539, at $z=8.6$ by \citet{Lehnert10} marks a milestone in observational cosmology that galaxies form less than 600 million years after the Big Bang. The detected $\lya$ line flux of this object is $\sim 6\times 10^{-18}\, \rm{ergs\, cm^{-2}\, s^{-1}}$, which implies a $\lya$ luminosity of $\sim 5.5 \times 10^{42}\, \rm{ergs\, s^{-1}}$. This is close to that of our galaxy at $z=8.5$. Our model suggests that this LAE may have a total mass of $\sim 10^{10-11}\, \Msun$, a stellar mass of $\sim 10^{9}\, \Msun$, a SFR of the order of $10 \;  \Msunyr$, and $f_{\rm esc}^{\lya} \sim {0.18}$. We note that \cite{Dayal11} suggested a lower SFR of $2.7-3.7\; \Msunyr$ for this galaxy by assuming a $\lya$ escape fraction of 100 {per cent}. However, our detailed RT calculations show that the escape fraction is much smaller than unity, which means it would require a higher SFR in order to produce the observed $\lya$ flux.

\subsubsection{The $\lya$ Line Profiles}

\begin{figure}
\begin{center}
\includegraphics[scale=0.4]{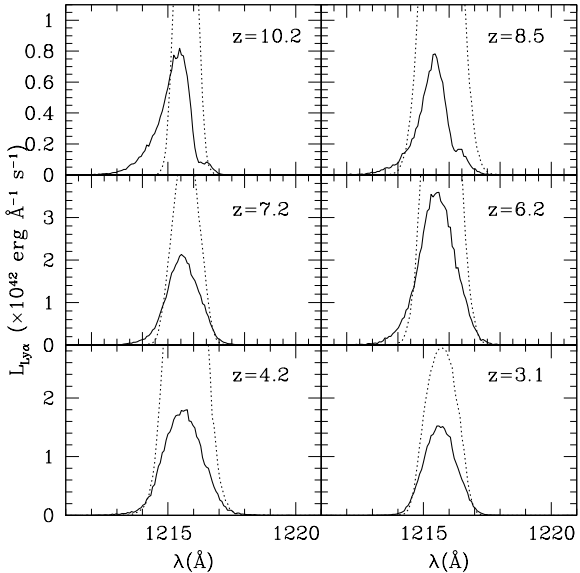}
\caption{ 
Emergent $\lya$ profile.
Dot and solid lines are the profiles of intrinsically emitted and
escaped photons respectively.
}
\label{fig:profile_cosmo}
\end{center}
\end{figure}

Figure~\ref{fig:profile_cosmo} shows the profiles of $\lya$ line of the modeled galaxies. 
The frequency of the intrinsic $\lya$ photon is sampled from a Maxwellian distribution with 
the gas temperature at the emission location in the rest flame of the gas. 
We collect all escaped $\lya$ photons and estimate the line profile.
In practice,  the inhomogeneous gas structure can cause the change of flux depending on the viewing angle.
Although the $\lya$ flux can change by some factors \citep{Yajima11b},
we focus on only the mean value in this paper. 

The $\lya$ lines of our galaxies show mostly a single peak, a common profile of LAEs at high redshifts \citep[e.g.,][]{Ouchi10}. In a static and optically thick medium, the profile can be double peaked, but when the effective optical depth is small due to high relative gas speed or ionization state, there might be only a single peak \citep{Zheng02}. In our case, the flow speed of gas is up to $\sim 300~\rm km~s^{-1}$, and the gas is highly ionized by stellar and AGN radiation, which result in a single peak. 

In addition, the profile at higher redshift shifts to shorter wavelength, and it shows the characteristic shape of gas inflow \citep{Zheng02}. Although our simulation includes feedback of stellar wind similar to that of \citet{Springel05d-2}, the $\lya$ line profile indicates gas inflow or symetric in these galaxies. Our result suggests that high-redshift star-forming galaxies may be fueled by efficient inflow of cold gas from the filaments. We will address this issue in detail in a forthcoming paper (Yajima et al. in preparation).

We note that there is a number of studies on inflow and outflow in $\lya$ lines, both observationally \citep{Kunth98} and theoretically \citep[e.g.,][]{Verhamme06-2, Dijkstra10, Dijkstra11, Barnes11}. It has been suggested that the IGM correction of 
$\lya$ emission may be enhanced due to inflow \citep[e.g.,][]{Dijkstra07} or decreased via outflows \citep[e.g.,][]{Verhamme06-2}. However, there is a large uncertainty in the prescription of wind feedback in current simulations. To investigate the velocity field of ISM in high-redshift galaxies, one needs simulations with a realistic wind model and a high spatial resolution comparable to that of line observations, which are currently unavailable and are beyond the scope of this paper. 
Moreover, we should point out that the line profile can change due to the difference of the sub-grid ISM models.

In addition, 
the line profile may be highly suppressed and changed by scattering in the intergalactic medium (IGM) \citep[e.g.,][]{Santos04, Dijkstra07-2, Zheng10, Laursen11}. 
The IGM effectively absorbs the $\lya$ photons at the line center and at shorter wavelengths by the Hubble flow \citep[e.g.,][]{Laursen11-2}. 
Therefore, the inflow feature in our profiles may be disappeared
and the shape may change to an asymmetric single peak with only photons at red wing. We will investigate the IGM correction in future work that includes propagation and scattering of $\lya$ and ionizing photons in the IGM.


\subsubsection{The Properties of Ionizing Photons}

\begin{figure}
\begin{center}
\includegraphics[scale=0.4]{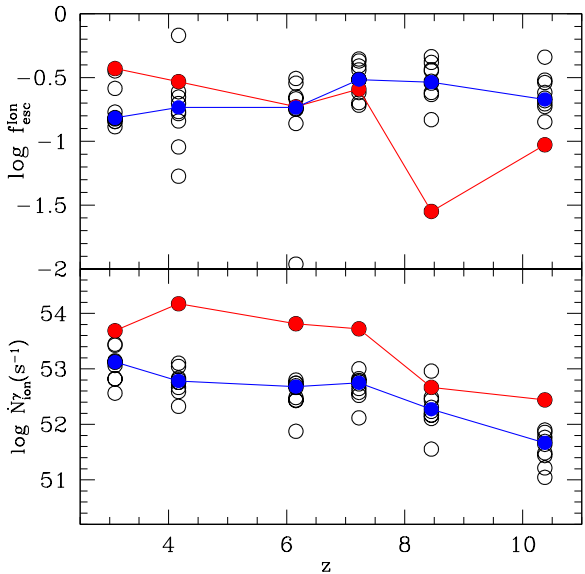}
\caption{The escape fraction of ionizing photons (top panel) and the resulting emissivity (bottom panel) of modeled galaxies at redshifts $z = 3.1 - 10.2$.
Red and Blue filled circles are the values of the most massive galaxies and the median in our sample at each snapshot, respectively.
}
\label{fig:fesc_ion}
\end{center}
\end{figure}

Figure~\ref{fig:fesc_ion} shows the escape fraction of ionizing photons, $f_{\rm esc}^{\rm Ion}$, and the resulting emissivity from both stars and AGNs of the modeled galaxies. 
The escape fraction of the most massive galaxies
has a range of  $f_{\rm esc}^{\rm Ion} \sim 0.03 - 0.37$, consistent with the results of \citet{Dijkstra11-2, Yajima11-2}. In addition, $f_{\rm esc}^{\rm Ion}$ decreases with increasing redshift. 
Unlike the $\lya$ and non-ionizing UV photons, the ionizing photons can be absorbed by hydrogen gas and dust. At high redshift, the star formation efficiency is small, so a large fraction of hydrogen gas remains neutral and prevent the escape of ionizing photons. 
The median values do not decrease with redshift due to lower dust mass.

The emissivity of the ionizing photons roughly increase with decreasing redshift, reaching $\sim {5\times 10^{53}}~\rm s^{-1}$ at $z=3.1$. in this work, we do not calculate the RT in the IGM, but the size of the H{\sc ii} bubble around the galaxy may be estimated by a modified Str\"{o}mgrem sphere. Assuming a clumpiness factor of  $C=5$ \citep{Iliev07, Pawlik09b}, the radius of the H{\sc ii} bubble is estimated to be $r = {554,~ 198,~47}$ and ${28}$ kpc at $z=3.1, ~6.2,~8.5$ and $10.2$, respectively. 
The $\lya$ photons may pass through the  H{\sc ii} bubbles, resulting in a correlation between high-redshift LAEs and  H{\sc ii} regions, as suggested by \cite{McQuinn07}.
In addition, the median value of ionizing-photons emissivity is smaller due to lower SFR, and hence the H{\sc ii} bubble becomes smaller than that of the most massive galaxy by a factor of $\sim 2-6$.

\section{Discussion}

In observations of $\lya$ emission from high-redshift galaxies, there are a number of  uncertainties regarding the contributions from AGNs \citep[e.g.,][]{Wang04, Ouchi08, Nilsson11}, stars, and excitation cooling from cold accretion. With our model, we can separate the radiation components and disentangle the contribution from different sources. Also, in numerical simulations, resolution plays an important role in the robustness of the results. So we discuss these physical and numerical effects on $\lya$ emission in this Section.

\subsection{Effects of AGN and Stellar Radiation on $\lya$ Emission}

\begin{figure}
\begin{center}
\includegraphics[scale=0.42]{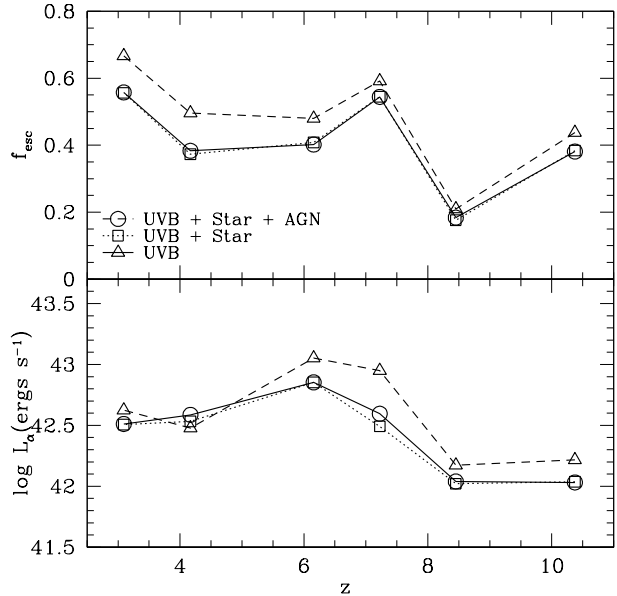}
\caption{Effects of AGN on escape fraction of $\lya$ photons (top panel) and the emergent luminosity (bottom panel). The ionization and $\lya$ luminosity are estimated with radiation from the following sources: UVB only  (triangles), UVB + star (squares), and UVB + AGN + star (circles).}
\label{fig:fesc_comp_star}
\end{center}
\end{figure}

The effects of AGN and stars on the $\lya$ emission is shown in Figure~\ref{fig:fesc_comp_star}. It appears that the presence of AGN does not change significantly the $\lya$ escape fraction. It is because AGNs are usually located in galactic center where the gas density peaks, 
most of the ionizing photons near Lyman-limit ($\sim 13.6~\rm eV$) are trapped there \citep[e.g.,][]{Nagamine10-2, Yajima11-2} and cannot escape. 
On the other hand, a fraction of high energy photons can escape from the galaxy without ionizing interstellar gas due to small cross section of gas and dust.
Therefore, AGN radiation has little effect on the $\fesc$ of $\lya$ photons. There is no change in $\fesc$ in the case where there is no stellar and AGN radiation.

In addition, the AGN does not affect the $\lya$ luminosity, as shown in the bottom panel of Figure~\ref{fig:fesc_comp_star}. This may be because the BHs in these galaxies are still quite small and the accretion rates are quite low. 
On the other hand, the presence of stars appears to change the escape fraction and emergent luminosity of $\lya$, because the stellar radiation is much stronger than that of AGN in the model galaxies, and the effect becomes stronger at lower redshift.  Because $\lya$ emissivity is produced by the recombination of ionizing photons and collisional excitation from hydrogen gas, the interplay between these two mechanisms determines the outcome. At redshifts $z > 6$, excitation cooling from cold accretion of dense gas becomes comparable to recombination cooling, while at $z \sim 3$, the number of $\lya$ photons by recombination in H{\sc ii} regions by stars and AGNs outnumbers that from excitation cooling. Therefore, the contribution from stars to the $\lya$ emission changes with cosmic time and environment.
The $\lya$ emissivity from excitation cooling can be reduced due to the stellar ionization \citep[e.g.,][]{Faucher10}. 
Moreover, the distribution of excitation $\lya$ cooling can be more extended than star-forming regions, resulting in higher escape fraction. 
Thus, the escape fraction and emergent luminosity of $\lya$ somewhat increase in the case without stellar radiation.

\subsection{Resolution Effect}

\begin{figure}
\begin{center}
\includegraphics[scale=0.42]{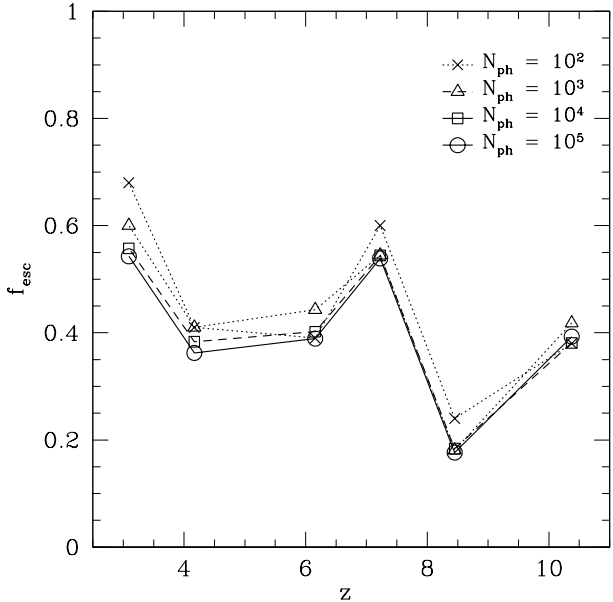}
\caption{
Effects of photon number on the escape fraction of Ly$\alpha$ photons.
Open triangles, squares and circles indicate photon packet number $N_{\rm ph} = {10^{2}}, 10^{3}, 10^{4}$, and $10^{5}$, respectively. The fiducial number used in our general RT calculations is $N_{\rm ph} = 10^{5}$.
}
\label{fig:fesc_comp_N}
\end{center}
\end{figure}

Figure~\ref{fig:fesc_comp_N} demonstrates effects of photon number on the escape fraction of $\lya$ photons, with different number of photon packets, $N_{\rm ph} = 10^{2}, 10^{3}, ~10^{4}$ and $10^{5}$. It appears that the results converge once the photon number is above $10^{4}$, which is an order of magnitude lower than the fiducial $N_{\rm ph} = 10^{5}$ in our general RT calculations. The relative difference between $N_{\rm ph} = 10^{4}$ and $10^{5}$ is 1-6 per cent.  
Of course, the photon number should increase with galaxy mass, but for our modeled galaxies, $N_{\rm ph} = 10^{5}$ should be sufficient. 

\begin{figure}
\begin{center}
\includegraphics[scale=0.42]{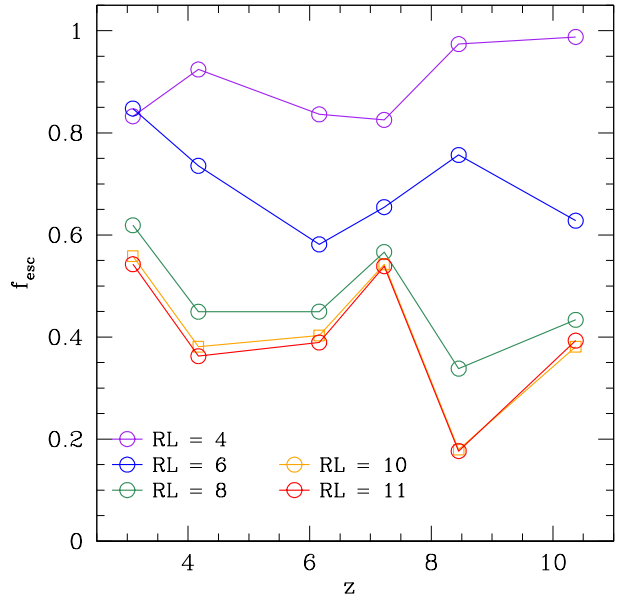}
\caption{
Effect of grid resolution on the escape fraction of Ly$\alpha$ photons.
Different color indicates different maximum refinement level (RL): 
purple -- RL = 4, blue -- RL = 6, green -- RL = 8, orange -- RL = 10, and red -- RL = 11. The fiducial RL used in our general RT calculations is 11. The cell size of the highest refinement level is $1/ 2^{\rm RL + 1}$ Mpc in comoving scale.
}
\label{fig:fesc_comp_L}
\end{center}
\end{figure}

The effects of grid resolution on the escape fraction of $\lya$ photons is presented in Figure~\ref{fig:fesc_comp_L}, where results from different maximum refinement level (RL) are compared. It appears that $\fesc$ changes dramatically with RL, but it converges at $\rm {RL} \ge 10$. Grids with poor resolution can produce artificially high $\fesc$ by a factor of more than 20. Because most of the $\lya$ photons are created in high-density gas clumps, where they are also effectively absorbed by numerous scatterings, sufficiently resolving the gas density field is important in the calculation of $\lya$ RT. If we follow the resolution of $\rm RL = 10$ with uniform grids, it would require a number of cells of $\sim 2048^{3}$, too expensive for RT simulations with the current computational facility. In fact, the current limit is $\sim 500^{3}$, as demonstrated by recent work on three-dimensional RT calculations using uniform grid \citep[e.g.,][]{Iliev07, Yajima11-2}. Therefore, adaptive refinement grid is critical in the RT calculation of $\lya$ emission.



%
%

\section{SUMMARY}

We have presented an improved, three-dimensional, Monte Carlo radiative transfer code $\art$ to study multi-wavelength properties of galaxies. It has the following essential features:

\begin{itemize}

\item It couples $\lya$ line, ionization of neutral hydrogen, and multi-wavelength continuum radiative transfer. The $\lya$ module includes emission from both recombination and collisional excitation, and the continuum includes emission from X-ray to radio.

\item It employs an adaptive grid scheme in 3-D Cartesian coordinates, which handles an arbitrary geometry and covers a large dynamic range over several orders of magnitude. Such a grid is critical in resolving the clumpy, highly inhomogeneous density field in the RT calculations. 

\item It adopts a two-phase ISM model in which the cold, dense clouds are embedded in a hot, diffuse medium in pressure equilibrium. In the case where hydrodynamic simulations have sufficient resolution to resolve the multi-phase ISM, this model ensures an appropriate prescription of the ISM physics, which is important for studying dust properties in galaxies.   

\item It includes a supernova-origin dust model in which the dust is produced by Type-II supernovae,
  which is especially relevant for dust in high-redshift, young objects, because there are insufficient low-mass, old AGB stars in these systems to produce the observed amount of dust from the classical dust models. 

\item It follows the radiation from stars and AGNs, and calculates the scattering, absorption, and re-emission of the surrounding medium. It efficiently and self-consistently produces SEDs and images in a full spectrum, including $\lya$ emission line and continuum from X-ray to millimeter, as well as ionization structures of the medium, for direct comparison with multi-band observations.

\item It can be easily applied to either grid- or particle-based hydrodynamical simulations
and has broad applications in cosmology.

\end{itemize}

We have tested the new $\art$ extensively, and found that the coupling of $\lya$ line and continuum enables a self-consistent and accurate calculation of the $\lya$ properties of galaxies, as the equivalent width of the $\lya$ line depends on the UV continuum, and the escape fraction of $\lya$ photons strongly depends on the ionization structure and the dust content of the object.

We applied the code to a cosmological SPH simulation of a Milky Way-like galaxy, and studied in details a sample of massive galaxies at redshift z=3.1 - 10.2. 
We find that most of these galaxies are $\lya$ emitters. 
The escape fraction of $\lya$ photons of the most massive galaxy in each snap shot increases with decreasing redshift, from about {0.18} at redshift $z \sim 8.5$ to {0.6} at $z \sim 3$. 
The emergent $\lya$ luminosity shows similar trend, increasing from $\La ={1.0} \times10^{42}\, \ergs$ at $z \sim 8.5$ to $\La = {2.3} \times10^{42}\, \ergs$ at $z \sim 3$. 
The equivalent widths of these galaxies change with redshift as well, from $\sim {67}\, \A$ at $z \sim 8.5$ to $\sim {21}\, \A$ at $z \sim 3$. 
The profiles of the resulting $\lya$ lines commonly show single peak, and the peak of the profile at $z \gtrsim 6$ is shifted to shorter wavelength, implying gas inflow from surrounding filaments. On the other hand, the median values in our galaxy sample shows somewhat different trend from that of the most massive ones. The escape fraction monotonically increases with redshift, while the $\lya$ luminosity shows a weak evolution, as a result of the interplay between SFR and escape fraction. 

Our results suggest that the LAEs and their properties evolve with cosmic time, and that the first LAEs in the early universe, such as currently the most distant one detected at $z \sim 8.6$ \citep{Lehnert10}, may be dwarf galaxies with a total mass of $\sim 10^{10-11}\, \Msun$ fueled by cold gas accretion. They may have a star formation rate of order of $10\, \Msunyr$,  and a $\lya$ escape fraction of $\sim 20$ per cent. 

%
%
\section*{Acknowledgments}
We thank Mark Dijkstra, Claude-Andr{\'e} Faucher-Gigu{\`e}re, Caryl Gronwall, Robin Ciardullo, Lars Hernquist, and Avi Loeb for stimulating discussions and helpful comments, as well as the referee for an insightful and constructive report which has helped improve the manuscript significantly. Support from NSF grants AST-0965694, AST-1009867 (to YL), and AST-0807075 (to TA) is gratefully acknowledged. YL thanks the Institute for Theory and Computation (ITC) at Harvard University where the project was started for warm hospitality. We acknowledge the Research Computing and Cyberinfrastructure unit of Information Technology Services at The Pennsylvania State University for providing computational resources and services that have contributed to the research results reported in this paper (URL: http://rcc.its.psu.edu). 
The Institute for Gravitation and the Cosmos is supported by the Eberly College of Science and the Office of the Senior Vice President for Research at the Pennsylvania State University.

%
%



\label{lastpage}

\end{document}